\documentclass[aps,prb,superscriptaddress,amsfonts,amsmath,amssymb,showpacs,floatfix,reprint,longbibliography,footinbib]{revtex4-2}

\usepackage{url}
\usepackage{bm,bbm}
\usepackage{graphicx}
\usepackage{color}
\usepackage[colorlinks=true, urlcolor=blue, linkcolor=blue, citecolor=blue, pdftex]{hyperref}
\usepackage[english]{babel}
\usepackage{physics}
\usepackage{slashed}

\usepackage{feynmp-auto}

\newcommand{\iu}{\mathrm{i}} 
\newcommand{\eu}{\mathrm{e}} 
\newcommand{\hc}{\mathrm{h.c.}} 
\newcommand{\bi}{\bm{i}}
\newcommand{\bj}{\bm{j}}
\newcommand{\bq}{\bm{q}}
\newcommand{\bk}{\bm{k}}
\newcommand{\bij}{\bi\bj}

\usepackage{soul}
\usepackage{blindtext}
\usepackage{lipsum}
\usepackage{nicefrac}

\newcommand{\dagga}{{\phantom{\dagger}}}

\newcommand{\todo}[1]{}
\newcommand{\diff}[1]{{#1}}

\begin{document}

\title{Dynamics and stability of U(1) spin liquids beyond mean-field theory:\\
Triangular-lattice $J_1$--$J_2$ Heisenberg model}

\author{Josef Willsher}
\thanks{\diff{willsher@pks.mpg.de}}
\affiliation{Technical University of Munich, TUM School of Natural Sciences, Physics Department, 85748 Garching, Germany}
\affiliation{Munich Center for Quantum Science and Technology (MCQST), Schellingstr. 4, 80799 München, Germany}
\affiliation{\diff{Max-Planck-Institut f\"{u}r Physik komplexer Systeme, 01187 Dresden, Germany}}

\author{Johannes Knolle}
\affiliation{Technical University of Munich, TUM School of Natural Sciences, Physics Department, 85748 Garching, Germany}
\affiliation{Munich Center for Quantum Science and Technology (MCQST), Schellingstr. 4, 80799 München, Germany}
\affiliation{Blackett Laboratory, Imperial College London, London SW7 2AZ, United Kingdom}

\date{\today}

\begin{abstract}
Quantum spin liquids (QSLs) are long-range entangled phases of frustrated magnets exhibiting fractionalised spin excitations.
In two dimensions, there is limited analytical understanding of their excitation spectra beyond parton mean-field theories, which fail to capture many features of the finite frequency dynamical response.
We use a self-consistent mean-field theory combined with the random phase approximation (RPA) for the $J_1$--$J_2$ Heisenberg model on the triangular lattice to describe the strong spinon-spinon interactions of the U(1) Dirac QSL.
We obtain quantitative results for the dynamical spin structure factor and phase diagram compatible with comprehensive numerical efforts.
We show the continuum response at the Brillouin zone corners is described by a spinon-exciton hybridised with monopole gauge excitations. We further show that the transition from the QSL to 120-degree coplanar order is in the QED$_3$ O(3)-GN universality class.
We extend the method to chiral QSLs and XXZ anisotropy, and discuss its broad range of applicability to other models and for describing inelastic neutron scattering experiments in KYbSe$_2$.
\end{abstract}

\maketitle

\section{Introduction}
Quantum spin liquids (QSLs) are spin-disordered phases of frustrated magnets which are characterised by their \emph{emergent gauge structure}, long-range entangled ground states and unconventional fractionalised `parton' excitations.
This includes spinons, fermionic particles carrying fractions of the spin flip quantum number of a magnon, as well as exotic emergent gauge excitations like `monopoles' or `visons'~\cite{wen2004quantum,fradkin2013field,savary2016,knolle2019}.
While QSLs are often classified by parton mean-field theory \cite{wen2002}, which identifies states by their gauge structure and universal low-energy properties, it has been much less successful at producing a quantitative theory of microscopic spin Hamiltonians.
In particular, the dynamical structure factor --- the quantity probed in inelastic neutron scattering experiments --- often contains collective excitations or finite-frequency sharp modes that cannot be captured by bare parton mean-field theory.
%

\begin{figure}
\centering
\includegraphics{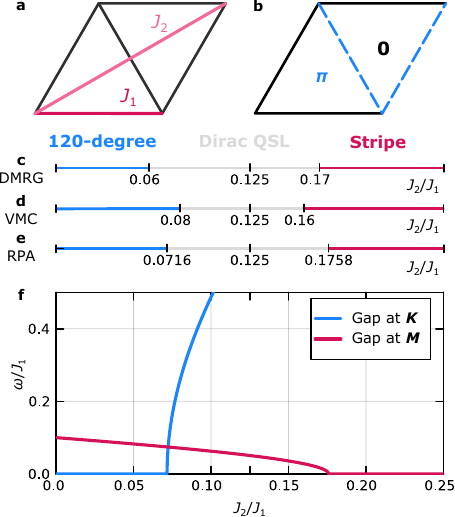}
\caption{
{Triangular lattice Heisenberg model and phase diagram}.
(a) Triangular lattice $J_1$--$J_2$ Heisenberg interactions along nearest and next-nearest neighbor bonds. (b) $\pi$-flux Ansatz shown on a two-site unit cell; black/colored bonds have $\pm$ sign in the Hamiltonian. 
(c,d) Phase diagram as a function of $J_2/J_1$ using density-matrix renormalization group (DMRG) \cite{hu2015} and variational Monte Carlo (VMC) \cite{iqbal2016} methods. The point $J_2/J_1=1/8$ is the classical 120-stripe transition point.
(e) Phase diagram evaluated using the random phase approximation (RPA) method of this work.
(e) Gap of competing spin excitations in the U(1) Dirac QSL.
\label{fig1}}
\end{figure}

Here, we provide an example where a parton theory is the starting point for a whole { \it quantitative} theory and obtain the rich excitation spectrum of a QSL.
We focus on a paradigmatic model in frustrated magnetism: the spin-half Heisenberg model on the triangular lattice,
\begin{equation}
    \mathcal{H}_{J_1 J_2} = \sum_{\langle\bm{i}\bm{j}\rangle} J_{\bm{i}\bm{j}}\, \vec{S}_{\bm{i}} \cdot \vec{S}_{\bm{j}} \, .
\label{eq:spin_hamiltonian}
\end{equation}
In the simplest case, $J_{\bm{i}\bm{j}}=J_1$ only on nearest neighbor bonds, we have the model where Anderson first proposed a QSL ground state in 1973 \cite{anderson1973}.
%
It was later shown that quantum fluctuations do not fully melt 120-degree order, but adding an AFM next-nearest-neighbor exchange $J_{\bm{i}\bm{j}}=J_2$, shown in Fig.~\ref{fig1}(a), can further frustrate the non-collinear ordering.
More recent numerical works \cite{kaneko2014,zhu2015,hu2015,iqbal2016,hu2019} have found a non-magnetic state around the maximally frustrated point $J_2/J_1=1/8$ \cite{jolicoeur1990,chubukov1992}.
%
In this context, numerical tools have made progress in determining the nature of the state: in particular, variational Monte Carlo (VMC) methods point to a very good variational wavefunction, given by the $\pi$-flux Ansatz. This zero-parameter variational state is described by
fermionic spinons hopping on a lattice with alternating 0 and $\pi$ fluxes, shown in Fig.~\ref{fig1}(b) \cite{iqbal2016}.
Because the mean-field theory has a low-energy Dirac dispersion and a compact U(1) gauge field, the resulting state is often called the U(1) Dirac QSL.
%
This picture has also received confirmation from DMRG calculations, with recent approaches finding a gapless state \cite{hu2019,drescher2023}, albeit this is still an active topic of discussion \cite{jiang2026,kovalska2026}.

Despite the success of this variational ansatz under Gutzwiller projection, a `zeroth-order' analytical approach to the problem, where this constraint is only imposed on average, is not able to describe any properties of the original spin model accurately.
Firstly, the nearest-neighbor mean-field Hamiltonian is entirely independent of the longer-range interaction $J_2$ that we know is necessary to stabilise the QSL. Furthermore, a standard Hartree--Fock treatment of the Hamiltonian always prefers gapped or ordered states.
Finally, we highlight that the dynamical structure factor without interactions or Gutzwiller projection \cite{ferrari2019} bears little resemblance to the interacting problem \cite{drescher2023}. That is, it lacks any low-energy features around the Brillouin zone corners, that was hypothesised by previous works to be due to fluctuations of the gauge field \cite{ferrari2019,drescher2023}.
%

\begin{figure}
\centering
\includegraphics{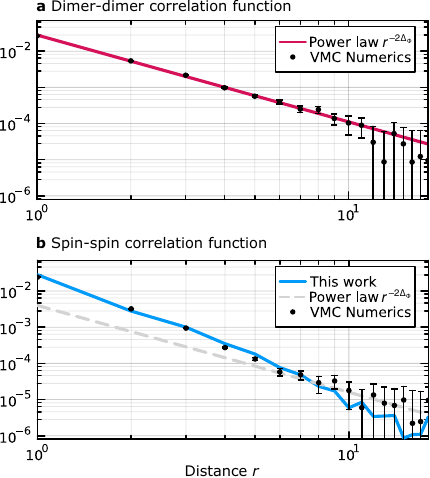}
\caption{
{Correlations of $J_1$--$J_2$ Heisenberg model: universal \emph{versus} non-universal behaviour}.
(a) Connected dimer-dimer correlation function $\langle {D}_{\bm{0}} {D}_{\bm{r}}\rangle-\langle D_0\rangle\langle D_{\bm{r}}\rangle$, $D_{\bm{r}} = \vec{S}_{\bi}\cdot\vec{S}_{\bj}$ with $\bm{ij}$ along nearest neighbor bonds and $\bm{r}$ connecting dimers in the perpendicular direction. Data calculated on a $L=42$ cluster, taken from Ref.~\cite{ferrari2025}. The data are well described by a power-law $r^{-2\Delta_\Phi}$ with $\Delta_\Phi\approx1.2$, signaling algebraic correlations of monopole operators.
(b) Spin-spin correlation function $\langle \vec{S}_{\bm{0}}\cdot \vec{S}_{\bm{r}}\rangle$
with $\bm{r}$ along next-nearest neighbor bonds.
We present VMC data (published in Ref.~\cite{willsher2025}), calculated on a $L=60$ cluster. 
The monopole power-law $r^{-2\Delta_\Phi}$ does not match the data, but instead we find good agreement with calculations presented in this work (we calculate the correlation function in the random-phase approximation including monopole-fermion coupling, $\alpha_{\mathcal{S}}=0$ and $g_{\mathcal{S}}=0.05$).
All data presented at $J_2=J_1/8$.
\label{fig2}}
\end{figure}

These recent numerical developments have been spurred also by parallel field-theoretical interest.
By coarse-graining the lattice model and focusing on the low-energy degrees of freedom only, one can gain more insight. The all-important single-occupancy constraint of the parton theory can be captured by coupling the fermions to a fluctuating U(1) gauge field \cite{wen2004quantum}.
In fact, it is predicted that the aforementioned $\pi$-flux lattice model realises a U(1) lattice gauge theory, which flows to a strongly coupled QED$_3$ conformal fixed point in the long-wavelength limit \cite{appelquist1988,karthik2019,song2019}. This is a critical phase with a large emergent $\mathrm{SU}(4)\times \mathrm{U}(1)$ global symmetry group.
The most relevant operators in this low-energy limit are so-called monopoles; these are instanton events which tunnel $2\pi$ flux of the emergent gauge field. Importantly, there are 6 of them which transform under lattice symmetries as order parameters of competing 120-degree Neel and valence bond solid order. For this reason, the U(1) Dirac spin liquid is often called the `mother of competing orders' \cite{hermele2005algebraic,song2019}.
Exact diagonalisation results have recently bolstered this picture even further, finding a large overlap between the low-lying eigenstates of the $J_1$--$J_2$ Heisenberg model in the highly frustrated regime with a family of zero-variational parameter wavefunctions of the QED$_3$ vacuum, combined with its monopole and fermion-bilinear excitations \cite{wietek2024}.
%
One central prediction of this framework is that the correlation functions of spin and dimer operators are algebraic, governed by a single critical exponent at long distances $\Delta$ \cite{hermele2005algebraic}.
We test this prediction in Fig.~\ref{fig2}, where we display numerical results for these correlation functions calculated using the VMC method from Refs.~\cite{ferrari2025,willsher2025}. While the dimer operator is well described by a power law $r^{-2\Delta}$ with $\Delta=1.25$, the spin correlation function is more compatible with a finite correlation length at odds with field theory predictions.
Despite the accumulation of evidence that the low-energy sector realises an emergent quantum electrodynamics, this discrepancy motivates the question of how non-universal features from the lattice-scale and finite frequency spin dynamics interplay with the low-energy physics.

It is not generally possible to predict which type of QSL is the ground state of a given spin Hamiltonian.
For this task we still rely almost exclusively on numerical methods, with the notable exception of the celebrated, exactly soluble Kitaev model \cite{kitaev2006anyons}.
%
%
%
In this work, we address this question for the triangular lattice $J_1$--$J_2$ Heisenberg model.
We predict the position and universality of magnetic ordering transitions analytically by incorporating fluctuations within the parton approach.
We thus develop methods to make quantitative predictions for the phase boundaries between magnetically ordered and QSL states. Our results are presented in Fig.~\ref{fig1}(e), showing strong agreement with numerical approaches. Later, we additionally produce phase diagrams for the $J_1$--$J_2$ model with additional spin-scalar chirality interactions and XXZ anisotropy, again comparing them favourably to numerical approaches.
We are additionally able to calculate dynamical correlation functions in the QSL phase. We leverage these results to interpret numerical findings, and understand more deeply the possible signatures of QSLs at finite-frequencies. 
%
%
%
Our method treats interactions between lattice fermions using the random phase approximation, as well as a perturbative treatment of the effects of low-energy gauge fluctuations due to monopoles.
%
%
On a broader conceptual level, our work develops the analytical tools to make quantitative predictions from a strongly-interacting quantum spin Hamiltonian, and demonstrates that there can be predictive power in parton approaches.

\section{Stability of QSL states to magnetic order}

We now show that the U(1) Dirac QSL state on the triangular lattice can predict instabilities to the correct competing orders as a function of $J_2/J_1$, at values compatible with numerical works.
To do this, we go beyond mean-field theory and include fluctuations of the competing magnetic order.
A common starting point for analytical treatment of spin Hamiltonians \eqref{eq:spin_hamiltonian} is a transformation into the Abrikosov (fermionic parton) representation
\begin{equation}\label{eq:abr}
    \vec{S}_{\bi} = \frac{1}{2} f_{\bi,\alpha}^\dagger\vec{\sigma}_{\alpha\beta}f^\dagga_{\bi,\beta} \, .
\end{equation}
This transforms the spin Hamiltonian Eq.~\eqref{eq:spin_hamiltonian} into an interacting four-fermion theory
\begin{equation}
        \mathcal{H}_{{J_1 J_2}} = -\frac{1}{2} \sum_{\bm{i}\bm{j}}
    J_{\bm{i}\bm{j}} \left(f_{\bm{i}\alpha}^\dagger f^\dagga_{\bm{j}\alpha}f_{\bm{j}\beta}^\dagger f_{\bm{i}\beta}^\dagga + \frac{1}{2} f_{\bm{i}\alpha}^\dagger f_{\bm{i}\alpha}^\dagga f_{\bm{j}\beta}^\dagger f_{\bm{j}\beta}^\dagga \right) \, ,
    \label{eq:interacting_H}
\end{equation}
up to a constant.
We approach the interacting fermion problem in a path-integral formulation \cite{weng1991}.
By performing a series of Hubbard--Stratonovich decouplings of \eqref{eq:interacting_H}, considering fluctuations of all the mean-field parameters, we find
\begin{multline}\label{eq:HS_full}
        \mathcal{H} = -\sum_{\bi\bj} J_{\bi\bj} \bigg [\frac{1}{2} \left( \chi_{\bi\bj} f^\dagger_{\bj} f^\dagga_{\bi} + \hc - |\chi_{\bi\bj}|^2\right) \\
    + \left( \vec{m}_{\bi} \cdot \vec{m}_{\bj} - \vec{m}_{\bi}\cdot(f^\dagger_{\bj}\vec{\sigma}f^\dagga_{\bj})\right)
    \bigg]
    \,.
\end{multline}
A mean-field theory of a QSL state is described by a saddle point $\chi_{\bi\bj} = \langle f^\dagger_{\bi\alpha} f_{\bj\alpha}^\dagga\rangle \neq 0$ with zero magnetic order $m_{\bi}^\alpha = 0$.
The excitations of this theory are free fermions, as well as a gauge boson field associated with phase rotations $f^\dagger_{\bm{i}\alpha} \to e^{i\theta_{\bm{i}}}f^\dagger_{\bm{i}\alpha} $ and $a_{\bm{i}\bm{j}} \to a_{\bm{i}\bm{j}} - \theta_{\bm{i}} + \theta_{\bm{j}}$.

Furthermore, there are additional \emph{amplitude fluctuations} of $m_{\bi}$ and $\chi_{\bi\bj}$.
These are often neglected in the typical treatment of parton theories, since they are gapped at the mean field level, but this is not guaranteed beyond the non-interacting limit.
Consider for example the amplitude fluctuations of the competing magnetic order $\vec{\phi}_{\bq}$ with ordering wavevector $\bq$.
We show in Appendix~\ref{app2} that the effective potential for amplitude fluctuations goes as $r_{\bq}^{\mathrm{MF}}|\phi_{\bq}|^2$ in the mean-field approximation.
For example, competing 120-degree order has a stiffness $r_{\mathrm{MF}}=(3J_1/2-3J_2)^{-1}$, which is positive for all $J_2/J_1<1/2$.
Beyond this mean-field approximation, we can integrate out the fermions to calculate a renormalised self energy correction to the Hubbard--Stratonovich mode
$r_{\bq}|\phi_{\bq}|^2 = [r_{\bq}^{\mathrm{MF}}-2\chi^0(\bq,\omega)]|\phi_{\bq}|^2$.
In order to interpret this self energy, we can expand the fermion susceptibility around $\omega=0$, where it is purely real $\Pi^0_{\bq} = \chi^0(\bq,\omega=0)$. We find a renormalised stiffness $r_{\bm{K}}=r_{\bm{K}}^{\mathrm{MF}}-2\Pi^0_{\bm{K}}$. 
which condenses when $\Pi^0_{\bm{K}} > r_{\bm{K}}^{\mathrm{MF}}/2$.
This allows us to solve for the critical coupling strengths to 120-degree magnetic ordering at $J_{2,c}^{120}/J_1 = 0.07158$ and stripe magnetic order at $J_{2,c}^{\mathrm{str}}/J_1 = 0.1758$.
%
Figure~\ref{fig1}(c)--(e) compares these values with numerical calculations using the DMRG and VMC methods \cite{hu2015,iqbal2016}, showing strong quantitative agreement.

\section{Dynamical response}

Our main interest is the spin susceptibility, where we focus on the diagonal ($zz$) component
\begin{equation}\label{eq:strucfac}
    \chi^{zz}(\bm{q},\omega) = \frac{1}{N}\sum_{\bi\bj} \eu^{-\iu \bm{q}\cdot (\bi-\bj)} \int \dd{t} \, \eu^{\iu\omega t} \langle{S}_{\bi}^z(t) {S}_{\bj}^z(0)\rangle\,,
\end{equation}
which we use to define the longitudinal spin structure factor $S^{zz}(\bm{q},\omega)=\operatorname{Im}\chi^{zz}(\bm{q},\omega)$
that is directly proportional to the cross section of inelastic neutron scattering (INS) experiments.
%
Given the MFT Eq.~\eqref{eq:H0fermions}, one can evaluate the free fermion-fermion susceptibility $[\chi^{0}(\bm{q},\omega)]^{\alpha\beta}$ but it fails to describe the strongly interacting model of Eq.~\eqref{eq:interacting_H}, e.g. there are no low-energy excitations around the $\bm{K}$ momenta at all.
We employ the RPA to treat the four-fermion interaction term \eqref{eq:interacting_H}, which consists of summing up an infinite series of bubble diagrams \cite{schrieffer1989,chubukov1992a}. 

\begin{figure}
\centering
\includegraphics{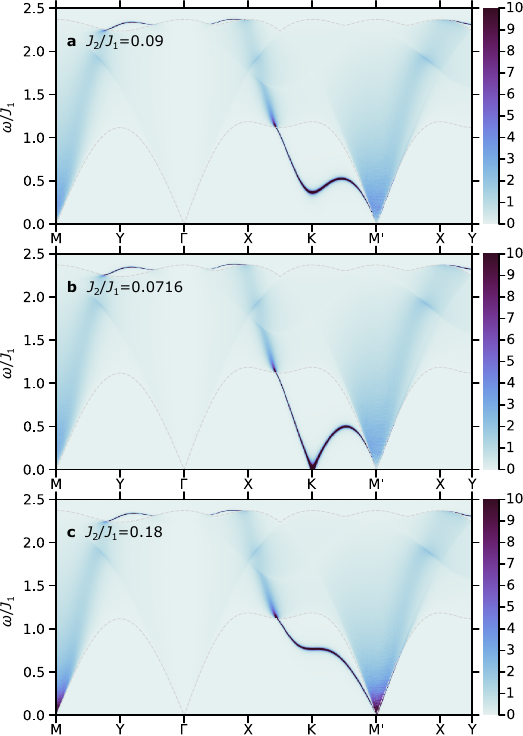}
\caption{{The dynamical structure factor $S^{zz}(\bm{q},\omega)=\operatorname{Im}\chi^{zz}(\bm{q},\omega)$}. Spinon susceptibility calculated in the random phase approximation with next-nearest neighbour coupling (a) $J_2/J_1 = 0.09$ in the U(1) Dirac QSL phase, (b) $J_{2}/J_1 = 0.0716$ at the transition to 120-degree order, and (c) $J_{2}/J_1 = 0.18$.
Gray dashed lines indicate the boundary of the two-spinon continuum; the sharp poles in the response are the spinon exciton.
Calculations performed with system size $L=720$.
\label{fig3}}
\end{figure}

Note that in itinerant {\it ordered} magnets described by Hubbard-like models similar self-consistent theories have  successfully described the dynamical spin response and collective spin wave modes~\cite{kampf1996collective,peres2002spin,singh2005,knolle2010,knolle2011multiorbital}, including on the triangular lattice~\cite{willsher2023}. 
Previous applications within parton MFT treated an on-site Hubbard $U$, which approximately imposes the on-site single-filling constraint \cite{ho2001,ma2018} or arises due to screened gauge fluctuations in a magnetised QSL with a Fermi surface \cite{balents2020}.
Here, we consider only the effect of interactions $[J(\bq)]_{XY}$ on dynamical response functions \cite{zhou2003,ghioldi2018dynamical,zhang2020}, given by the real-space nearest- and next-nearest-neighbor interaction $J_{\bi\bj}$, as Eq.~\eqref{eq:interacting_H}. The RPA susceptibility in terms of the interaction reads:
\begin{equation}
    {\chi}(\bm{q},\omega) = \left[1 +2{J}(\bq)\cdot {\chi}^{0}(\bm{q},\omega) \right]^{-1} \cdot {
    {\chi}^{0}(\bm{q},\omega)}\, ;
    \label{eq:rpa_chi}
\end{equation}
we emphasise that all quantities here are matrices with sublattice indices, derived in Supplementary Note 1. 
The elements of the interaction matrix ${J}(\bq)$ are given in terms of $J_{1,2}$ and the bandwidth of the bare ${\chi}^{0}$ is fixed by the MFT QSL order $t/J_1 = 0.198$, such that there are no adjustable parameters in our calculation.
%
%
The spinon contribution to the dynamical structure factor is shown in Fig.~\ref{fig3}(a) with $J_2/J_1=0.09$.
The most striking effect of the spinon interactions is to create a sharp collective mode coming down to low energy; this \emph{spinon exciton} is a quasiparticle with quantum numbers of a magnon arising as a spinon particle-hole bound state.
This quasiparticle emerges as a pole in $\chi(\bq,\omega)$, or equivalently a zero in $[1+2J\chi^0(\bq,\omega)]$ from Eq.~\eqref{eq:rpa_chi}. This is undamped outside of the two-spinon continuum as $\chi^0$ is purely real.
It displays a sharp magnetoroton-like minimum around $\bm{K}$, a signature of the proximate 120-degree order.
The mode broadens when entering the two-spinon continuum (highlighted with gray dashed lines), and exits above it again around the $\bm{\Gamma}$ point.
The only broad continuum at low energies is present at the $\bm{M}$ points from inter-Dirac cone transitions.
At the $\bm{M}$ point, the mode is at low energies within the spinon continuum, leading to an enhancement of low-energy spectral weight relative to the free-fermion MFT case (c.f. Fig.~\ref{figapp}(a) in Appendix~\ref{app1}). 
The spectral weight remains weaker than the sharp mode around $\bm{K}$, in agreement with time-dependent DMRG results~\cite{drescher2023}.
Overall, we find that the sharp spinon exciton dominates the response in the fermion parton approach using self-consistent parton mean-field theory plus RPA.
All of these findings are in remarkable agreement with VMC results \cite{ferrari2019} and DMRG studies of the stable U(1) Dirac QSL regime \cite{drescher2023,drescher2025}. We include in the Supplementary Note 2 a detailed comparison to these data sets. 
Our analytical approach complements these `black-box' methods by providing an origin of the observed signatures, and we will later use this as a foundation to study additional spin interactions and including gauge fluctuations.

Reducing the ratio $J_2/J_1$ causes the mode to lower at the $\bm{K}$ point, until at $J_2/J_1=0.0716$ it goes to zero energy, shown in Fig.~\ref{fig3}(b). This signals an instability of the QSL phase to 120-degree magnetic order and coincides with the results of our Hubbard--Stratonovich approach.
%
%
%
%
Expanding the fermion susceptibility around zero frequency, we derive the action of this critical boson
\begin{equation}\label{eq:critical_boson}
    S_{\phi} = \sum_{\bq}  \left[\omega^2 + c^2(\bq-\bm{K})^2 + \alpha^2 r_{\bm{K}}\right]|\vec{\phi}_{\bq}|^2
\end{equation}
with a mass defined $m^2 = \alpha^2 r_{\bm{K}}$ for a constant $\alpha$ and velocity $c$, see Appendix~\ref{app2} for details.
We plot this excitation gap as a function of $J_2/J_1$ in Fig.~\ref{fig1}(e).
The quadratic theory has a three-component relativistic soft mode (note we have neglected gauge fluctuations so far) with dynamical exponent $z=1$. 
The condensed bosons continuously become the Goldstone modes of the 120-degree ordered state, with ordering wavevector $\bm{K}$.
%
%
Tuning $J_2>0.18\, J_1$, the pole of Eq.~\eqref{eq:rpa_chi} also diverges at $\bm{M}$ as $\omega\to0$, signaling a tendency for stripe magnetic order. 
The structure factor is shown in Fig.~\ref{fig3}(c)
and the mass at $\bm{M}$ likewise plotted in Fig.~\ref{fig1}(e).
We will now briefly argue that this transition should be first order in nature.
Condensing magnetic order at $\bm{M}$ in a continuous transition would lead to the spontaneous formation of a mass at the Dirac cones; due to consequent confinement of monopoles, this would then tend towards the proliferation of 120-degree order, which is energetically disfavoured for $J_2/J_1 > 1/8$.
Instead, we view it as more likely that the transition point is preempted by a first-order transition to stripe magnetic order. This is compatible with numerous numerical works, which find a first-order transition around $J_2 /J_1 \approx 0.15$ \cite{kaneko2014,hu2015,zhu2015,iqbal2016,hu2019,ferrari2019,drescher2023}).


\section{Gauge fluctuations}

What is the effect of gauge field fluctuations on the dynamical response of the QSLs?
Previous works have studied the stability and excitations of the U(1) Dirac QSL in first-order mean-field theory \cite{song2019}. 
It predicts that the low-energy response around the $\bm{K}$ points are dominated by \emph{spin-triplet monopole} excitations.
The structure factor diverges with a quantum-critical behaviour
\begin{multline}\label{eq:monopower}
    \chi_{\mathcal{S}}(\bm{q}, \omega+\iu \eta) = \alpha_{\mathcal{S}}^2\langle \mathcal{S}^\dagger \mathcal{S}\rangle(\bq,\omega)\\ = {\frac{c \alpha_{\mathcal{S}}^2}{(|\bm{K}-\bm{q}|^2-\omega^2 -\iu \eta)^{3/2-\Delta_\Phi}}} \, ,
\end{multline}
as evaluated in Appendix~\ref{app1}.
At the point $\bq=\bm{K}$, this simplifies to $\chi_{\mathcal{S}}(\bm{K}, \omega) = {c \alpha_{\mathcal{S}}^2} \, \omega^{2\Delta_\Phi-3}$.
Here,
$\Delta_\Phi$ is the monopole scaling dimension, and
$\alpha_{\mathcal{S}}$ is a numerical constant that defines the weight of the spin-triplet $S=1$ monopole operators in the continuum field theory with lattice-spin operators, as discussed in Appendix~\ref{app1}.
This constant cannot be fixed by symmetry arguments, meaning it is \textit{a priori} unclear how much intensity monopoles contribute to correlation functions.

We can constrain the monopole contribution by comparing this prediction with numerical calculations.
First of all, let us make a general qualitative observation:
We can integrate out the gapped boson at energies below the spinon exciton gap and recover the continuum QED$_3$ action. Here, the long-distance spin correlations must be algebraic.
However, the contribution of monopoles to the spin structure factor must be small in VMC and DMRG on finite size lattices, since no monopole continuum is observed below the spinon exciton in the dynamical structure factor \cite{ferrari2019,drescher2023,drescher2025}.
To quantify this, we return to the real-space spin-spin correlations $g(r) = \langle\vec{S}_{r}\cdot\vec{S}_0\rangle$ in the direction of next-nearest-neighbor bonds, shown in Fig.~\ref{fig2}(b).
The dashed line shows an upper bound on the power-law contribution that would arise from algebraic real-space monopole correlations.
%
In this way, we constrain the constant to be $\alpha_{\mathcal{S}}^2 \leq 0.004$.
That is, the matrix element of monopoles is very small or zero.

\begin{figure*}
\centering
\includegraphics{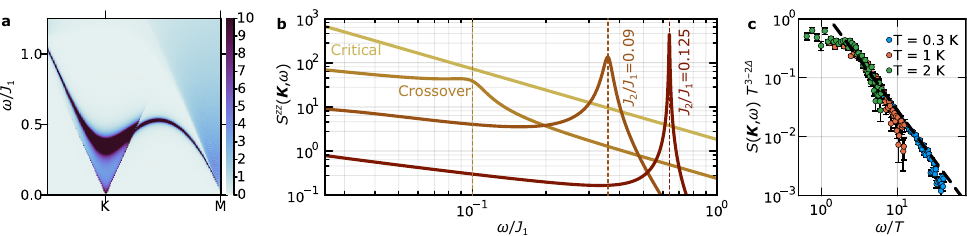}
\caption{
{Hybridisation of the spinon exciton with critical monopole fluctuations and experimental detection of QED$_3$ O(3)-Gross--Neveu (GN) exponents.}
(a) Momentum-resolved structure factor $S^{zz}(\bq,\omega)$ with a spinon exciton-monopole coupling $g=0.05$. We assume the monopole matrix element is zero, and all continua are due to self-energy corrections of the collective mode.
(b) Spectral weight at the $\bm{K}$ point for several values of $J_2/J_1$. 
The results at $J_2/J_1 = 0.125$ and $0.09$ are evaluated in the modified RPA scheme (see SM). The low-energy behaviour is a power law $\sim \omega^{2\Delta_\Phi-3}$ with QED$_3$ exponents.
At the critical point (spinon exciton mass $m^2\to 0$) the low-energy behaviour is a modified power law $\sim \omega^{2\Delta_\Phi^{\mathrm{GN}}-3}$ with QED$_3$ O(3)-GN exponents.
In the crossover regime ($m=0.1 J_1$) the structure factor interpolates the two power laws in the regimes $\omega \ll m$ and $\omega\gg m$.
(c) Scaling collapse of the divergent dynamical structure factor $S(\bm{K},\omega)$ in KYbSe$_2$, reproduced from Ref.~\cite{scheie2024proximate}. Dashed line shows the `critical' scaling $\sim \omega^{2\Delta_\Phi^{\mathrm{GN}}-3}$ with QED$_3$ O(3)-GN exponents.
\label{fig4}}
\end{figure*}

%
Although we have shown that the monopole's direct contribution to spin-spin correlators is vanishingly small, it is possible for monopoles to modify the spectral response through their indirect coupling to the spinons.
Now, we will take this coupling into account on the field-theoretical level, and use it to make predictions of signatures of gauge fluctuations, as well as the nature of the phase transition.
The monopole and fluctuating spinon exciton have the same quantum numbers, and will therefore generically couple as
\begin{equation}\label{eq:phiPhi_inter}
        S_{\phi\mathcal{S}} = g_{\mathcal{S}} \int\dd[3]{x} \,  (\phi_\alpha^*(x)\mathcal{S}_\alpha + \hc) \, .
\end{equation}
The total action is given by $S_{\mathrm{QED}_3} + S_{\phi} + S_{\phi\mathcal{S}}$, where the boson action is written in
Eq.~\eqref{eq:critical_boson}.
First, let us consider the perturbative limit.
We find that the coupling leads to a hybridization of the spinon exciton and quantum-critical fluctuations of the spin-1 monopoles \cite{ghaemi2006}.
For small coupling $g_{\mathcal{S}}$, we may assume that this coupling does not destroy the spin liquid state, and evaluate a self-energy correction to the spinon exciton due to monopole fluctuations; in Appendix~\ref{app2}, we derive
\begin{equation}\label{eq:critical_boson_monopole}
    S_{\phi}^{\mathrm{eff}} = \sum_{\bm{q}}[\omega^2 + c^2(\bq-\bm{K})^2 + \alpha^2 r - g^2_{\mathcal{S}} \langle \mathcal{S}^\dagger \mathcal{S}\rangle(\bq,\omega)]|\vec{\phi}_{\bq}|^2
\end{equation}
using conformal perturbation theory.
Identifying the monopole susceptibility in Eq.~\eqref{eq:monopower}, we can see that this perturbative correction gives an imaginary component to the self-energy of the spinon exciton, which diverges as $m^{2\Delta_\Phi-3}$ as the gap $m=\alpha \sqrt{r}$ goes to zero.
We calculate in Appendix~\ref{app2} that this leads to a broadening of the spinon exciton mode.
This phenomenon is similar to the hybridization of spin-0 monopoles and phonons recently discussed in Refs.~\cite{seifert2024,lee2019}.
We confirm this prediction of the spinon exciton behaviour obtained from our continuum field theory with a lattice-scale RPA calculation in Fig.~\ref{fig4}(a). 
In this approach, we treat a coupling between the monopole and the fermion spin $g_{\mathcal{S}} = 0.05$, which introduces new diagrams into our RPA series with intermediate monopole propagators, as described in Supplementary Note 3. 
%
The resulting effect is a broadening of the spinon exciton mode, and the appearance of a critical monopole continuum at low frequencies, even when taking $\alpha_{\mathcal{S}} = 0$.
We present the resulting real-space correlation as the blue curve in Fig.~\ref{fig2}(b), which fits all VMC data points.
%
%

Finally, we would like to understand the nature of the spinon exciton condensation transition in the presence of a dynamical gauge field.
As we approach the transition, the perturbative treatment breaks down and the spinon exciton fully hybridises with the monopole continuum.
To proceed, we instead return to Eq.~\eqref{eq:phiPhi_inter} and integrate out the fluctuating boson. Through this procedure, we generate a retarded monopole-monopole interaction
\begin{equation}\label{eq:sphiphi}
    S_{\Phi\Phi} = - g^2_{\mathcal{S}}\int\dd[3]{x}\dd[3]{y} \mathcal{S}^\dagger_\alpha(x)G_\phi(x-y)\mathcal{S}^\dagga_\alpha(y) \, ,
\end{equation}
where $G_\phi(x-y)$ is the propagator of the spinon exciton.
For momenta $q^2 \ll \alpha^2 r$, this interaction is short ranged and can be expanded $-(g^2_{\mathcal{S}}/\alpha^2 r)\mathcal{S}^\dagger_\alpha(x)\mathcal{S}^\dagga_\alpha(x+\epsilon) +\cdots$, which can then be expanded in terms of charge-0 primary operators using the operator product expansion as $\epsilon\to0$.
Although the operator $\mathcal{S}^\dagger_\alpha\mathcal{S}^\dagga_\alpha$ is irrelevant, the diverging prefactor $-g^2_{\mathcal{S}}/\alpha^2 r$ leads to a new unstable fixed point.
In Appendix~\ref{app2}, we argue that the dominant OPE channel comes from 4-fermion interactions of the form $(\overline\psi\vec{\sigma}\psi)^2-(\overline\psi\vec{\tau}\psi)^2$, resulting in a transition in the universality class QED$_3$ O(3)-Gross--Neveu (GN) \cite{dupuis2019}
(also known as QED$_3$ chiral Heisenberg GN).
At this point, the symmetry group $\mathrm{SU}(4)\simeq \mathrm{SO}(6)/\mathbb{Z}_2$ of QED$_3$ is broken explicitly to $\mathrm{SO}(3)\times \mathrm{SO}(3)$.
This broken symmetry splits the scaling dimensions of the spin-triplet and spin-singlet monopole operators; large-$N_f$ results suggest that the scaling dimension of spin triplet monopoles is lowered $\Delta_{\Phi}^{\mathrm{GN}} \approx  0.78 <\Delta_\Phi$ \cite{dupuis2019}.
This picture should be confirmed with a renormalisation group calculation of the critical point.
Importantly, monopoles are fully hybridised with the critical boson at this point, and the monopoles must hence contribute strongly to the dynamical spin structure factor (irrespective of the small matrix element $\alpha_{\mathcal{S}}$ in the QED$_3$-only theory).
We expect a divergent continuum controlled by the monopole scaling dimensions at the new conformal fixed point; specifically $\chi_{\mathcal{S}}(\bm{K}, \omega) \sim \omega^{2\Delta_\Phi^{\mathrm{GN}}-3}$.
%

Away from the exact critical point, we expect an appropriate scaling collapse in terms of $\mathcal{G}(\omega/m)$, as described in Supplementary Note 3.
Approaching the critical point, the correlation length will diverge $\xi\sim m^{-1}$, and the real-space spin-spin correlations  will cross over to QED$_3$ O(3)-GN exponents $g(r)\sim r^{-2\Delta^{\mathrm{GN}}_\Phi}$. 
We summarise in Fig.~\ref{fig4}(b) the equivalent result in momentum-space. Here, we observe the broadening of the spinon exciton as $m$ is decreased which then crosses over into a critical regime where QED$_3$ O(3)-GN exponents dominate.
A key signature of critical spin liquids in experiments will be the identification of power-law divergences in the spectral function, and its appropriate temperature scaling collapse \cite{hermele2008}.
We predict that at the critical point between a U(1) Dirac QSL and a 120-degree ordered state, plotting $T^{3-2\Delta} S(\bm{K},\omega;T)$ against $\omega/T$ produces a collapse and directly probes the scaling dimension of monopole operators $\Delta$ at the QED$_3$ O(3)-GN critical point.
This analysis was performed on KYbSe$_2$ \cite{scheie2024proximate}, shown in Fig.~\ref{fig4}(c), which is a triangular lattice compound with ground state 120-degree order and a large frustration ratio due to a $J_2/J_1$ just below the transition to a spin-disordered state.
The authors found a scaling collapse with exponent $1.73(12)$ \cite{scheie2024proximate},
which is compatible with a monopole scaling dimension $\Delta=0.64(12)$.
These critical exponents are compatible with the quantum-critical scaling expected near the QED$_3$ O(3)-GN transition to a U(1) Dirac QSL, suggesting there might already exist an experimental realization of this critical point from the quantum critical fan of the weakly ordered material in the ytterbium delafossite family.

\section{Additional interactions}

\begin{figure}
\centering
\includegraphics{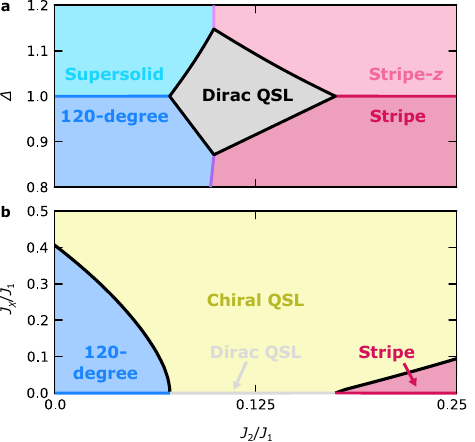}
\caption{{Phase diagram of the $J_1$--$J_2$ Heisenberg model in the presence of perturbations}. (a) XXZ anisotropy $\Delta$ reduces the stability of the QSL and leads to new supersolid Y and stripe-z orders. (b) Spin scalar chirality interaction $J_\chi$ stabilises the chiral spin liquid.
\label{fig5}}
\end{figure}

In this Section, we include additional interactions and use it to calculate the phase diagram and dynamical structure factor of the resulting QSLs away from the symmetric Heisenberg limit.
We study the spin liquid states in the presence of two perturbations: XXZ anisotropy $\Delta$ which breaks SU(2) spin symmetry, and spin-scalar chirality $J_\chi$, which breaks time-reversal symmetry, hence gapping the spinons and introducing higher-order fermion interactions.
Figure~\ref{fig5} shows a summary of the resulting phase diagrams, which is in good agreement with recent numerical studies \cite{gallegos2025,wietek2017,saadatmand2017,gong2017}.

\subsection{XXZ anisotropy}
We consider an anisotropic spin interaction,
\begin{equation}\label{eq:xxz}
    \mathcal{H}_{\mathrm{XXZ}} = \sum_{\langle\bm{i}\bm{j}\rangle} J_{\bm{i}\bm{j}}\, \left( S^x_{\bi}S^x_{\bj} + S^y_{\bi}S^y_{\bj} +\Delta S^z_{\bi}S^z_{\bj} \right) \, ,
\end{equation}
where we break SU(2) spin-rotation symmetry down to U(1) for $\Delta\neq1$ \cite{zhu2017,gallegos2025}.
In Appendix~\ref{app3} we show that the self-consistent mean-field Hamiltonian is also given by an SU(2) symmetric pi-flux state, with a hopping now given by $t_{\bij}=J_1(1+\Delta)\chi_{\bij}/4$.
Although the Hamiltonian and bare fermion susceptibility preserve SU(2) symmetry, the interaction matrix $J$ now takes the form $\operatorname{diag}(1,1,\Delta)$ in spin-component space. For this reason, we compute the general spin susceptibility matrix $\chi^{ij}(\bq,\omega)$ and focus on the $\chi^{zz}$ and $\chi^{\pm}$ components, both of which contribute to the symmetric sum $\chi^{ii}$.

\begin{figure}
\centering
\includegraphics{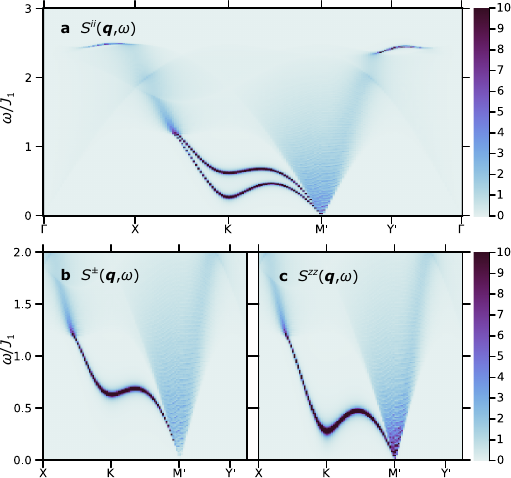}
\caption{{Structure factor of the U(1) Dirac QSL with XXZ anisotropy}. We show (a) the full diagonal components $\sum_iS^{ii}(\bq,\omega)$ with $S^{ij}(\bm{q},\omega)=\operatorname{Im}\chi^{ij}(\bm{q},\omega)$ as well as (b) transverse $S^\pm(\bq,\omega)$ and (c) longitudinal components $S^{zz}(\bq,\omega)$. We take next-nearest neighbor exchange $J_2/J_1 = 0.1$ and XXZ anisotropy $\Delta=1.1$; calculations performed with system size $L=240$.
\label{figxxz}}
\end{figure}

The dynamical structure factor is shown in Fig.~\ref{figxxz} for easy-axis anisotropy $\Delta=1.1$, $J_2/J_1=0.1$.
We observe a splitting of the spinon exciton into two non-degenerate modes, with fluctuations in the $S^{zz}$ component having a lower gap.
Because the Ansatz is SU(2) symmetric, the static structure factor is approximately isotropic in the various channels, but real-space correlations may have different correlation lengths. 
This explains the apparently isotropic correlations observed using DMRG \cite{gallegos2025} at distances smaller than the correlation length.

Increasing anisotropy $\Delta > 1$ eventually causes this mode to condense; for low $J_2$ this occurs at the $\bm{K}$ point, leading to supersolid order with in-plane 120-degree and out-of-plane Ising order.
For large $J_2$ the resulting phase is a so-called stripe-$z$ order. The resulting phase diagram for general $\Delta$ is shown in Fig.~\ref{fig5}(a).
The resulting condensed phases agree with DMRG investigations of the model \cite{gallegos2025}, though RPA seems to underestimate the stability of the liquid phase.
The nature of the phase transitions is modified by the lowered spin symmetry of the Hamiltonian. For \diff{easy-axis anisotropy $\Delta > 1$} the in-plane \diff{fermion bilinear} condenses and the resulting transition is QED$_3$ O(2)-GN (also known as chiral-XY GN) \cite{dupuis2022}.
This provides a mechanism for condensing supersolid order in the XXZ model, which has been proposed as a way to describe gapped Dirac continua in Na$_2$BaCo(PO$_4$)$_2$ \cite{jia2024}.
Alternatively for \diff{easy-plane anisotropy $\Delta<1$}, the transition is QED$_3$ $\mathbb{Z}_2$-GN (chiral Ising) universality class since only one \diff{fermion bilinear} is condensed.
The different symmetries lead to different monopole scaling dimensions and hence different power-law divergences in the different spin channels $S^{zz}(\bm{K},\omega)$ and  $S^{\pm}(\bm{K},\omega)$. 
%

\subsection{Spin scalar chirality}
Next, we perturb the $J_1$--$J_2$ model with a scalar spin chirality (SSC) term~\cite{hu2016variational} that breaks time-reversal symmetry
\begin{equation}\label{eq:ssc}
    \mathcal{H} = \mathcal{H}_{J_1J_2} + J_\chi\! \sum_{\bi\bj\bk \in \triangle,\bigtriangledown} \!\!\! \vec{S}_{\bi}\cdot (\vec{S}_{\bj}\cross \vec{S}_{\bk})\, ,
\end{equation}
where $\bi\bj\bk$ is an ordered set of the three spins on each triangular plaquette.
The SSC term induces six-fermion interactions; decoupling them in the mean-field channel induces a staggered phase in the mean-field state \eqref{eq:H0fermions}, shifting it from $[\pi,0]$ to $[\pi-\phi, \phi]$, where we derive $\phi \sim \chi\, J_\chi / J_1$ in Appendix~\ref{app4} \cite{banerjee2023,maity2024}.
We find that SSC interactions induce a (singlet) fermion mass term linear in $J_\chi$ \cite{hu2016variational}.
The fermion bands acquire a non-zero Chern number and can be integrated out at low-energies, leading to a Chern--Simons effective theory for the gauge fluctuations \cite{kalmeyer1987,wen1989}. 
The resulting state is a chiral spin liquid state with explicitly broken time-reversal symmetry.

We now evaluate the excitations in the random phase approximation.
The SSC induces six-fermion interactions, which cannot be treated simply within the RPA. 
Working in the limit of small $J_\chi\ll J_1$, we show that the dominant contribution of the SSC at order $(J_\chi/J_1)^2$ can be captured into effective four-fermion terms across nearest and next-nearest neighbor bonds in Supplementary Note 4.
Fig.~\ref{figchiral} displays the resulting spin structure factor with $J_\chi/J_1=0.1$ and $J_2/J_1=0.2$.
We see clearly that a gap opens in the spectrum at $\bm{M}$, such that the spinon exciton emerges beneath the fermion continuum, enhancing the spectral weight here.
We find excellent agreement with the DMRG calculations of Ref.~\cite{drescher2025}, for $J_\chi/J_1=0.1$ and $0.2$, $J_2/J_1=1/8$ too, shown in Supplementary Note 2.
A second effect of this gap opening is that it reduces the fermion susceptibility at zero-frequency, $\Pi^0_{\bm{M}}$ and hence increases the self energy of the spinon exciton; the result is that we predict an increase in the stability of the chiral spin liquid to stripe ordering under increasing $J_\chi$.
On the other side, we find that the effective decrease in nearest-neighbor interaction $J_1$ raises the energy of the spinon bound state at $\bm{K}$. This acts to lower the critical coupling $J_{2,c}$ until at $J_\chi/J_1\approx 0.4$, the CSL becomes stabilised in the nearest-neighbor only model $J_2=0$.
We display the phase diagram for low-$J_\chi$ in Fig.~\ref{fig5}(b).
We find very good agreement with the position of these phase boundaries found using $L=36$ exact-diagonalization and variational methods \cite{wietek2017,saadatmand2017,gong2017}.
Our method misses the proposed tetrahedral state
\cite{cookmeyer2021four}, a non-colinear state with magnetic order at the $\bm{M}$ point.
We suggest that this order cannot be precipitated through a second-order transition via the condensation of fluctuating magnetic order, since $J_\chi$ acts to stabilise the spinon exciton at this wavevector (note this is also seen in DMRG \cite{drescher2025}). 
%

\begin{figure}
\centering
\includegraphics{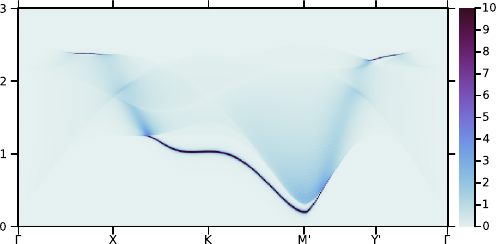}
\caption{{Structure factor of the chiral QSL} $S^{zz}(\bm{q},\omega)=\operatorname{Im}\chi^{zz}(\bm{q},\omega)$. We take next-nearest neighbour exchange $J_2/J_1 = 0.2$ and spin-scalar chirality
$J_\chi/J_1 =0.1$. 
All energies $\omega$ and interactions $J_2, J_\chi$ in units of $J_1$; calculations performed with system size $L=480$.
\label{figchiral}}
\end{figure}

\section{Discussion}

The hardest theoretical challenges obstructing the path to discovering a QSL are our limited ability to predict their experimental signatures, as well as the difficulty in predicting their stability. In other words, it is not clear what to look for or where to look for them \cite{knolle2019}.
%
%
%
%
We used a self-consistent parton mean-field theory plus fluctuations treated in the random-phase approximation to study the dynamical response and stability of the U(1) QSL on the triangular lattice for realistic spin Hamiltonians. 
We found that the spin dynamics is dominated by a collective spin fluctuation due to an emergent spinon exciton. This manifests as a magnon-like quasiparticle pole at finite frequency in the dynamical structure factor. 
Given that the U(1) Dirac QSL can be considered a melted 120-degree ordered state, we confirmed that a proper treatment of the lattice-scale interactions makes the finite-frequency response resemble that of the ordered state closely.
We evaluate the dynamical response and phase diagrams of extended triangular lattice Hamiltonians, including spin scalar chirality or XXZ anisotropy.
We confirm strong agreement between our results for the dynamical spin-structure factor and recent state-of-the-art numerical calculations in Supplementary Note 2, suggesting that the parton approach has a strong predictive potential for triangular lattice QSLs.
%

We investigated the effect of U(1) gauge fluctuations by including a coupling between fluctuating magnetic order and emergent monopoles.
This explains how algebraic correlation functions emerge at low energies from a lattice model, and is compatible with the picture of hybridisation
between spinon-bilinear and monopoles recently highlighted in numerical works \cite{budaraju2025}, where it was noted that fermion bilinear excited state wavefunctions have a large overlap with monopole states.
Furthermore, we predicted the critical scaling of the dynamical structure factor and identify the universality class of the transition to magnetic order as QED$_3$ O(3)-GN.
This ties into a promising avenue towards confirming the existence of fractionalised QSL states, inspired by studies of 1D Heisenberg spin chains \cite{lake2005}.
In these compounds, there is a consensus that the dynamics at finite energies is controlled by emergent Luttinger liquid physics, even though the materials order at low-temperatures due to weak 3D inter-chain coupling \cite{mourigal2013fractional}.
In light of recent evidence for (almost) gapless QSL-like response in triangular lattice magnets, i.e. YbZn$_2$GaO$_5$~\cite{bag2024evidence} and in the Yb-delaffossites~\cite{scheie2024proximate,scheie2024b}, we highlight the importance of understanding the critical properties of gapless QSL states and their transitions to neighbouring orders.
In the case of KYbSe$_2$, the value of $J_2/J_1$ is close to the 120-degree to spin liquid transition and we showed in Fig.~\ref{fig4}(c) that the low-energy continuum permits a scaling collapse compatible with critical exponents for the QED$_3$ O(3)-GN critical endpoint of the extended QSL phase.
We hope our method will allow a concrete comparison of model predictions to INS experiments in order to pin down microscopic Hamiltonians, for example quantifying the magnitude of XXZ anisotropy.

On the methodological side there are many exciting future directions.
Instabilities to $\mathbb{Z}_2$ QSLs can be understood by studying the susceptibility in the fermion pairing channel, and the condensation of a spin-hopping mode might also account for the QSL to valence bond solid (VBS) transition seen in the square-lattice $J_1$--$J_2$ model~\cite{morita2015quantum}. 
We also think a great insight could be gained by calculating other dynamical correlation functions, for example the dynamical dimer-dimer correlator. Numerical results of the U(1) DSL show that static dimer-dimer correlations are dominated by monopoles~\cite{ferrari2025}.
Given the complicating influence of the low-lying pole at the $\bm{K}$ point on spin-1 correlation functions, we suggest that these spin-singlet correlations might be a more promising route for identifying a U(1) Dirac QSL.

A famous example of a simple parton state is the Laughlin wavefunction which has provided the basis for our understanding of charge fractionalisation in strongly interacting electron gases \cite{laughlin1983}.
However, there is the prevailing belief that the parton approach applied to QSLs is less accurate and lacks  quantitative predictive power, for example in connection with doped superconductors \cite{brinckmann1999,brinckmann2001}.
In this work, we have shown that the $\pi$-flux wavefunction for extended triangular lattice Heisenberg models sets itself apart by providing quantitative results for dynamical correlations and magnetic phase diagrams.
We interpret the surprising success of the RPA method in this context as being compatible with the overwhelming picture also emerging from the numerical literature: that the $J_1$--$J_2$ AFM Heisenberg Hamiltonian has a ground state that is very well described by a wavefunction with zero variational parameters \cite{iqbal2016}. This is further bolstered by ED calculations which find the excitations are described by emergent QED$_3$ \cite{wietek2024}.
%
We suggest a promising line of future research: 
If a mean-field QSL parton state is a good variational Ansatz for a magnetic Hamiltonian (for example, as established by VMC or ED), the RPA approach is a simple yet powerful tool for describing the dynamical response and calculating the phase diagram.
This is corroborated by a complementary study investigating the $\mathbb{Z}_2$ Kitaev QSLs~\cite{rao2025}, where the ground state is known exactly.
Identifying further Hamiltonians which meet these criteria will advance the field beyond the study of fine-tuned models towards experimental realization.

\vspace{1.5cm}

\paragraph*{Acknowledgments} 
We would like to thank C.D. Batista, F. Becca, M. Drescher, F. Ferrari, HuiKe Jin, M. Knap, F. Pollmann, P. Rao, U.F.P. Seifert and O.A. Starykh for insightful discussions. We thank A. Scheie and A. Tennant for helpful discussions regarding the experimental relevance of our work. We furthermore thank F. Becca and F. Ferrari for allowing us to reproduce data from Ref.~\cite{ferrari2019}.
We also thank T. Bader, R. Eto, \diff{Y. Iqbal}, U.F.P. Seifert for helping to identify errors in previous versions of this manuscript, as well as L. Walther and R. Moessner for ongoing related collaboration.
J.K. also thanks for the hospitality of the Aspen Center for Physics, which is supported by National Science Foundation grant PHY-2210452. J.K. acknowledges support from the Deutsche Forschungsgemeinschaft (DFG, German Research Foundation) under Germany’s Excellence Strategy (EXC–2111–390814868 and ct.qmat EXC-2147-390858490), and DFG Grants No. KN1254/1-2, KN1254/2-1 TRR 360 -- 492547816 and SFB 1143 (project-id 247310070), as well as the Munich Quantum Valley, which is supported by the Bavarian state government with funds from the Hightech Agenda Bayern Plus. J.K. further acknowledges support from the Imperial-TUM flagship partnership. 



\appendix

\section{Parton mean field theory: $\pi$-flux Ansatz}\label{app1}

\subsection{Zeroth order mean-field theory}
The generic Heisenberg Hamiltonian in the Abrikosov fermion representation reads 
\begin{align}
        \mathcal{H}_{{J_1 J_2}}
        &= \frac{1}{4}\sum_{\langle\bm{i}\bm{j}\rangle} \left(f^\dagger_{\bi\alpha}\vec{\sigma}_{\alpha\beta}f^\dagga_{\bi\beta}\right) \cdot \left(f^\dagger_{\bj\alpha'}\vec{\sigma}_{\alpha'\beta'}f^\dagga_{\bj\beta'}\right)\\
        &= -\frac{1}{2} \sum_{\bm{i}\bm{j}}
    J_{\bm{i}\bm{j}} \left(f_{\bm{i}\alpha}^\dagger f^\dagga_{\bm{j}\alpha}f_{\bm{j}\beta}^\dagger f_{\bm{i}\beta}^\dagga + \frac{1}{2} f_{\bm{i}\alpha}^\dagger f_{\bm{i}\alpha}^\dagga f_{\bm{j}\beta}^\dagger f_{\bm{j}\beta}^\dagga - f_{\bm{i}\alpha}^\dagger f_{\bm{i}\alpha}^\dagga \right) \, .
    \label{eq:interacting_H_app}
\end{align}
In the triangular lattice case, we take the mean-field decoupling $\chi_{\bij} = \langle f^\dagger_{\bm{i}\alpha}f_{\bm{j}\alpha}^\dagga \rangle$ to find
\begin{equation}\label{eq:H0fermions}
{\mathcal{H}}_{\rm 0} 
=-\frac{J_1}{2}\sum_{ \bm{i}\bm{j}} (\chi_{\bm{i}\bm{j}} f^\dagger_{\bm{j}\alpha} f^\dagga_{\bm{i}\alpha} + {\rm h.c.} )
= -\sum_{ \bm{i}\bm{j}} (t_{\bm{i}\bm{j}} f^\dagger_{\bm{j}\alpha} f^\dagga_{\bm{i}\alpha} + {\rm h.c.} )
\end{equation}
where we define the hopping $ t_{\bm{i}\bm{j}} = J_{\bi\bj} \langle f^\dagger_{\bm{i}\alpha} f^\dagga_{\bm{j}\alpha} \rangle/2$. This has complex nearest-neighbour hoppings with uniform amplitude, which we write as $t_{\bm{i}\bm{j}} = t \mathrm{e}^{\mathrm{i} a_{\bm{i}\bm{j}}}$.
We also enforce the single-occupancy constraint on average $\langle f^\dagger_{\bm{i}\alpha} f^\dagga_{\bm{i}\alpha} \rangle = 1$.
We focus on the U(1) Dirac spin liquid in this work, formed by a $\pi$-flux Ansatz $t_{\bm{i}\bm{j}} = \pm t$ with alternating $0$ and $\pi$ fluxes through adjacent triangular plaquettes, see Fig.~\ref{fig1}(b)~\cite{iqbal2016}.

We calculate the dynamical spin structure factor Eq.~\eqref{eq:strucfac} from the main text. Taking the Fourier transform, we write the susceptibility as
\begin{equation}\label{eq:suscgen}
    \chi^{zz}(\bm{q},\iu\omega)
    = \frac{1}{N} \int \dd{\tau} \, \eu^{\iu\omega \tau} \langle T_\tau \,  {S}_{\bm{q}}^z(\tau) {S}_{-\bm{q}}^z(0)\rangle\, .
\end{equation}
Then substituting the Abrikosov fermion representation \eqref{eq:abr}, and contracting external fermions in the free-fermion approximation, we evaluate the leading contribution from the bare fermion bubble
\begin{fmffile}{freebubble_approx}
\begin{equation}
[\chi^{0}(\bm{q},\omega)] =\,\,
\frac{1}{2}\sigma^z\,
\begin{gathered}
\begin{fmfgraph*}(33,30)
 \fmfleftn{l}{1}\fmfrightn{r}{1}
 \fmf{fermion,right=.5}{l1,r1}\fmf{fermion,right=.5}{r1,l1}
\end{fmfgraph*}
\end{gathered}\,
\frac{1}{2}\sigma^z
\label{eq:free_bubble_baredef}
\end{equation}
\end{fmffile}\noindent
This result in Matsubara frequencies can then be analytically continued to real frequencies in the retarded scheme $\iu \omega \to \omega + \iu \eta$.
All calculations of fermion susceptibility are presented with a broadening given by $\eta = 0.8 \, J_1 / L$, where we scale it with the linear system size $L$.
We evaluate this retarded susceptibility in Supplementary Note 1, and plot the result in Fig.~\ref{figapp}(a).
There are low-energy excitations at the $\bm{M}$ point that correspond to particle-hole excitations across Dirac cones.

To approximately account for the local-moment renormalization induced by Gutzwiller projection, we rescale the Hamiltonian by a factor of two. In the unprojected mean-field state, $\langle\vec{S}_{\bi}^2\rangle_{\mathrm{MFT}}=3/8$, whereas projection onto single occupancy restores the physical value $\langle\vec{S}_{\bi}^2\rangle_{\mathrm{G}}=3/4$. This enhancement is modeled by $\vec{S}_{\bi}\rightarrow\sqrt{2}\vec{S}_{\bi}$, implying $\vec{S}_{\bi}\cdot\vec{S}_{\bj}\rightarrow2\vec{S}_{\bi}\cdot\vec{S}_{\bj}$ or equivalently $J_{\bij}\to2J_{\bij}$. For comparison, the two-spin problem gives a renormalization factor $8/5\simeq1.6$ from the unprojected singlet–triplet gap $5J/8$, rather than the physical gap $J$. 
We approximately account for this effect by rescaling the overall Hamiltonian with a phenomenological factor of two by rescaling $J_{\bij}\to 2J_{\bij}$. 
This is applied to both the self-consistent mean-field, as well as the interaction vertex (introduced below).

\begin{figure*}
\centering
\includegraphics{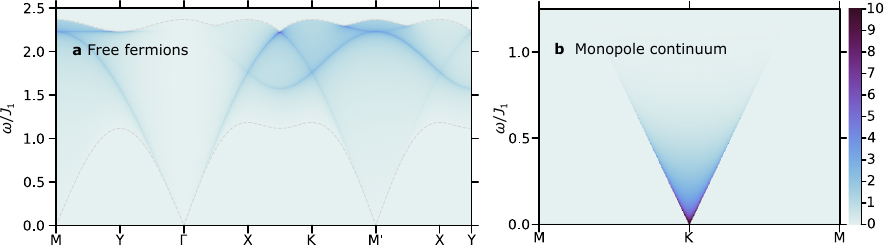}
\caption{Contribution to the dynamical structure factor $S(\bq,\omega)$ (a) in the free fermion approximation (b) considering only the monopole contribution.
All energies $\omega$ in units of $J_1$.
Gray dashed lines indicate the boundary of the two-spinon continuum in (a); there is finite spectral weight inside this whole region away from $\bm{q}=\bm{\Gamma}$ and gapless Dirac cone excitations at $\bq=\bm{M}$.
In the monopole calculation (b), the susceptibility diverges as $\omega^{2\Delta_\Phi-3}$ at $\bq=\bm{K}.$
\label{figapp}}
\end{figure*}

\subsection{First order mean-field theory}
The model has a U(1) gauge redundancy $f^\dagger_{\bm{i}\alpha} \to e^{i\theta_{\bm{i}}}f^\dagger_{\bm{i}\alpha} $ and $a_{\bm{i}\bm{j}} \to a_{\bm{i}\bm{j}} - \theta_{\bm{i}} + \theta_{\bm{j}}$, which implies that the gauge excitations are gapless.
In the parlance of Ref.~\cite{wen2002}, `first-order' MFT accounts for the effect of these gapless transverse (phase) fluctuations of the singlet order parameters on the stability of the QSL Ansatz.
The resulting low-energy field theory has $N_f=4$ Dirac cones described by the continuum action in Euclidean time \cite{kim1999,rantner2001}
\begin{equation}\label{eq:qed3action}
  \mathcal{L}_{\mathrm{QED}_3} = - \sum_{i=1}^4 \overline{\psi}_i \, \iu \gamma^\mu (\partial_\mu - \iu a_{\mu})\,  \psi_{i} + \frac{1}{4e^2} f_{\mu\nu}^{\ 2}\, .
\end{equation}
In the continuum, this theory is believed to flow to a conformally invariant, strongly coupled fixed point with an emergent $\mathrm{SU}(4)\times \mathrm{U}(1)_{\mathrm{topo}}$ symmetry \cite{appelquist1988,karthik2019}.
Since our gauge theory emerged from a lattice model, it must be compact. The most relevant gauge field fluctuations are monopole instanton events $\Phi(x)$, which tunnel $2\pi$ emergent flux~\cite{hermele2005algebraic}.
The symmetry properties of these monopoles, and hence the fate of the phase, is highly lattice-dependent; in the case of the triangular lattice, no relevant monopoles exist which are singlets under the lattice symmetry group, and so the QSL is stable~\cite{song2019,song2020}.
The monopoles $\Phi_a(x)$ form a 6-dimensional representation of $\mathrm{SU}(4)\simeq \mathrm{SO}(6)/\mathbb{Z}_2$ with $a=1\dots6$, which can be decomposed into $\mathrm{SO}(6)\simeq \mathrm{SO}(3)_{\mathrm{valley}}\times \mathrm{SO}(3)_{\mathrm{spin}}$.
Therefore, we split $\Phi_a = (\mathcal{S}_\alpha,\mathcal{D}_\alpha)$ into spin-triplet monopoles $\mathcal{S}_\alpha$ and valley-triplet $\mathcal{D}_\alpha$ monopoles, with $\alpha = 1\dots3$.
All monopoles have the same scaling dimension $\Delta_\Phi$ in the QED$_3$ theory, which can be approximated by $\Delta_\Phi \approx 1.02$ in a large-$N_f$ approach \cite{pufu2014,dyer2013}, and within the range $1.18$-$1.34$ from numerical estimates \cite{karthik2019}.
The correlation functions of monopoles are completely fixed as a power law; when the continuum theory holds, we have
\begin{equation}\label{eq:monopole_corrfn_cft_app}
    \langle \Phi^\dagger(x)\Phi(y)\rangle \sim |x-y|^{-2\Delta_\Phi} \, ,
\end{equation}
where $x=(\tau,\bm{x})$ is a coordinate in Euclidean spacetime.

Focusing on the spin structure factor at the $\bm{K}$ point, we expect a contribution from the spin-triplet monopoles which have this momentum \cite{song2019}.
The symmetry transformation properties of the monopoles tell us that they transform like the spin operator
\begin{equation}
    S_{\bi}^\alpha \simeq \operatorname{Re}[\iu \mathcal{S}^\alpha \eu^{\iu \bm{r}_{\bi} \cdot \bm{K}}] \, .
\end{equation}
We may introduce a numerical constant to quantify this; in momentum space we write $S_{\bm{K}}^\alpha = \alpha_{\mathcal{S}} \mathcal{S}^\alpha$, where $\alpha_{\mathcal{S}} $ is undetermined by the symmetry arguments.
This immediately tells us, via Eq.~\eqref{eq:monopole_corrfn_cft_app}, that spin-spin correlation functions along appropriate lattice directions must be power laws $\sim\alpha_{\mathcal{S}}^{\ 2} r^{-2\Delta_\Phi}$.
%
We can further substitute this identity into the definition of the spin structure-factor \eqref{eq:strucfac} to derive their contribution to the dynamical response.
We derive in Supplementary Note 3
\begin{fmffile}{monopole-1}
\begin{align}\label{eq:monosusc_th}
    \chi_{\mathcal{S}}(\bm{q}, \omega+\iu \eta) 
    &= \alpha_{\mathcal{S}}^2 \langle \mathcal{S}^\dagger \mathcal{S}\rangle (\bq,\omega+\iu \eta) \\
     &= {\frac{c\alpha_{\mathcal{S}}^2}{(|\bm{K}-\bm{q}|^2-\omega^2 -\iu \eta)^{3/2-\Delta_\Phi}}}
\\
&=
\,\,
S^z\,
\begin{gathered}
\begin{fmfgraph*}(50,10)
    \fmfleft{i1}
    \fmfright{o1}
    \fmf{boson}{i1,o1}
\end{fmfgraph*}
\end{gathered}
\, S^z \, .
\end{align}
\end{fmffile}\noindent
We stress, however, that this monopole continuum is only the expected limiting response at lowest energies, fixed by the constant $\alpha_{\mathcal{S}}$, and for a given lattice model the scale below which it appears can be small.
Thus, it may not describe the dominant non-universal response at higher energy as relevant for experiments and finite size numerics. 

\section{Paramagnon}\label{app2}

\subsection{RPA}
The effect of the interaction on the spin susceptibility can be accounted for by resuming a series of bubble diagrams. The interacting spin susceptibility \eqref{eq:suscgen} can be expanded in terms of the bare susceptibility as follows
\begin{fmffile}{susceptibilityseries}
\begin{equation}\label{eq:susceptibilityseries_diagrams_three}
\begin{gathered}
\begin{fmfgraph*}(60,30)
 \fmfleftn{l}{1}\fmfrightn{r}{1}\fmfrpolyn{shaded}{G}{4}
 \fmf{fermion,right=0.25,tension=2}{l1,G1}\fmf{fermion,right=0.25,tension=2}{G2,l1}
 \fmf{fermion,right=0.25,tension=2}{r1,G3}\fmf{fermion,right=0.25,tension=2}{G4,r1}
\end{fmfgraph*}
\end{gathered}
\,\,
=
\,\,
\begin{gathered}
\begin{fmfgraph*}(33,30)
 \fmfleftn{l}{1}\fmfrightn{r}{1}
 \fmf{fermion,right=.5}{l1,r1}\fmf{fermion,right=.5}{r1,l1}
\end{fmfgraph*}
\end{gathered}
\,\,
+
\,\,
\begin{gathered}
\begin{fmfgraph*}(80,30)
 \fmfleftn{l}{1}\fmfrightn{r}{1}
  \fmfdot{K1}\fmfdot{K2}
 \fmf{fermion,fermion,right=.5}{l1,K1}\fmf{fermion,fermion,right=.5}{K1,l1}
 \fmf{dashes,tension=4}{K1,K2}
 \fmf{fermion,fermion,right=.5}{r1,K2}\fmf{fermion,fermion,right=.5}{K2,r1}
\end{fmfgraph*}
\end{gathered}
\,\,
+\cdots
\end{equation}
\end{fmffile}\noindent
or 
\begin{equation}\label{eq:apprpa}
    \chi(\bm{q},\omega) = \chi^{0}(\bm{q},\omega)+ \chi^{0}(\bm{q},\omega) U_{\bq} \chi^{0}(\bm{q},\omega) + \cdots \, .
\end{equation}
where the interaction is the momentum space Heisenberg interaction \eqref{eq:interacting_H_app}.
The RPA naturally neglects diagrams not of this bubble form, such as vertex corrections.
We solve this equation by resumming the infinite series of bubbles to find
\begin{equation}
    \chi(\bm{q},\omega)= [1- U_{\bq} \chi^{0}(\bm{q},\omega)]^{-1} \chi^{0}(\bm{q},\omega) \, .
\end{equation}
The $S^z$ component of the Heisenberg interaction contributes to this series of diagrams, written as $J_{\bq} \,S^z_{\bq} S^z_{-\bq}$.
A perturbative calculation in Supplementary Note 1 defines the interaction matrix 
\begin{equation}
    U_{\bq}  = -2J_{\bq} \, .
\end{equation}
This RPA equation \eqref{eq:apprpa} must be defined in the 2-site unit cell, which we do in the Supplementary materials.
We find a collective mode, called a spinon exciton, which is a particle-hole bound state of spinons. Its energy is defined by the pole of the RPA condition $\det[1+2J_{\bq}  \chi^{0}(\bm{q},\omega)]=0$.
This excitation has the quantum numbers of a magnon in a spin-disordered state, as such we will also refer to it as a paramagnon.
Note, since the bare MFT is only dependent on $J_1$, the $J_2$-dependence of the state only enters through the RPA interaction vertex.
The Gutzwiller rescaling of the Hamiltonian leaves the RPA pole condition unchanged, since taking $t\to2t$ rescales $\chi^0\to (1/2)\chi^0$ and $J_{\bq}\to2J_{\bq}$.

\subsection{Continuum action}
The results show a dominant pole at low energies outside the bare spinon continuum, controlled by the interaction strength $J_2$.
We can attempt to describe this bosonic mode analytically by performing a Hubbard--Stratonovich transformation to decouple the spin-spin interactions 
in terms of the bond-pairing $\chi_{\bi\bj}$ and spin channel $\vec{m}_{\bi}$ simultaneously,
\begin{align}
        {\cal H} &= \sum_{\langle\bm{i}\bm{j}\rangle} J_{\bm{i}\bm{j}}\, \vec{S}_{\bm{i}} \cdot \vec{S}_{\bm{j}} 
    \\
    \nonumber
    &= \frac{1}{2}\sum_{\bi\bj} J_{\bi\bj} \Bigg [-\frac{1}{2}\left(\chi_{\bi\bj}f^\dagger_{\bj\alpha}f^\dagga_{\bi\alpha} + \hc- |\chi_{\bi\bj}|^2 \right) 
     \\ &\quad\quad\quad\quad+ \vec{m}_{\bi}\cdot \left(f^\dagger_{\bj\alpha}\vec{\sigma}_{\alpha\beta}f^\dagga_{\bj\beta}\right) - \vec{m}_{\bi}\cdot \vec{m}_{\bj}
    \Bigg],
    \label{eq:HS_full_app}
\end{align}
as written in the main text Eq.~\eqref{eq:HS_full}.
We decouple in both channels with equal weight $H= (1/2) H_\chi+ (1/2) H_m$. The mean field QSL Ansatz is defined by the simultaneous imposition of the following saddle-point equations 
\begin{align}
    \chi_{\bi\bj} &= \langle f^\dagger_{\bi\alpha}f^\dagga_{\bj\alpha}\rangle 
    \label{eq:mfconstr_chiij}
    \\
    \vec{m}_{\bi} &= \frac{1}{2}\langle f^\dagger_{\bi\alpha}\vec{\sigma}_{\alpha\beta}f^\dagga_{\bi\beta} \rangle = 0 \, .\label{eq:mfconstr_mi0}
\end{align}
Note that we must additionally impose the single-occupancy condition \cite{wen2002}
\begin{equation}
    \langle f^\dagger_{\bi\alpha}f^\dagga_{\bi\alpha}\rangle = 1\,,\quad
    \langle f^\dagger_{\bi\alpha}f^\dagga_{\bi\beta}\rangle \epsilon^{\alpha\beta} = 0 \, .
\end{equation}
As was done in the lattice calculation above, we rescale the interaction with a factor of two to account for the renormalization of energy scales provided by Gutzwiller projection (once again this only changes the overall normalization of energy scales, not the phase boundaries).

We proceed by defining $t_{\bi\bj}=J_{\bi\bj}\chi_{\bi\bj}/2$ and $\vec{\phi}_{\bi}=\sum_{\bj}J_{\bi\bj}\vec{m}_{\bj}$ \cite{bose2025b}.
In the spin sector this is equivalent to Weiss exchange fields.
The hopping sector becomes $t_{\bi\bj}f^\dagger_{\bj}f^\dagga_{\bi} + \hc$, which we take to be the Dirac Ansatz with a self-consistent hopping magnitude.
In this appendix, we will henceforth focus on the magnetic sector (c.f. Refs. \cite{bose2025,bose2025b} where both contributions are considered in the ordered phases with relative phenomenological weights)
and evaluate the gap of the fluctuating magnetic order $\vec{m}_{\bi}$.
The momentum-space action for this Weiss field is
\begin{equation}\label{eq:app_hs_spin_inter}
      \mathcal{H} [\vec{\phi}_{\bq}] =\sum_{\bq} 
      \left( -J_{\bq}^{\ -1} \vec{\phi}_{-\bq}\cdot\vec{\phi}_{\bq} + 2\vec{S}_{-\bq}\cdot\vec{\phi}_{\bq} \right),
\end{equation}
where $S^i_{\bq} = \frac{1}{2}\sum_{\bm{k}} f^\dagger_{\bk+\bq,\alpha}\sigma^i_{\alpha\beta} f^\dagga_{\bk,\beta}$.
Firstly, let us confirm that the amplitude fluctuations of spin order are gapped in the mean field;
ignoring the fermion-magnetisation coupling, we find
\begin{equation}\label{eq:hs_h_fluc}
  \mathcal{H}_{\mathrm{MF, free}}[\vec{\phi}_{\bq}] 
    =\sum_{q} m_0^{\, 2}(\bq) \, |\vec{\phi}_{\bq}|^2  \, ,
\end{equation}
where the mass is $m_0^{\,2}(\bq) = -J_{\bq}^{-1}$. Evaluating for the high-symmetry points, we find
\begin{equation}\label{eq:m0_q_km}
  m_0^{\,2}(\bm{K}) = \frac{2/3}{J_1-2J_2} \, , \quad
  m_0^{\,2}(\bm{M}) = \frac{1}{J_1+J_2}\, .
\end{equation}
The mean field gaps are indeed finite for $-1 < J_2/J_1 < 1/2$.

Corrections to the effective potential Eq.~\eqref{eq:hs_h_fluc} beyond the mean-field approximation is provided by the spinon-boson interaction written in Eq.~\eqref{eq:app_hs_spin_inter}.
We evaluate the resulting effective action to leading order when integrating out the spinons \cite{weng1991,scholle2024,bose2025b}:
\begin{equation}\label{eq:app_hsbosonenergy}
  \mathcal{H}[\varphi_{\bq}] 
    =\sum_{q} \left[ m_0^{\, 2}(\bq) - 2\chi^0(\bq,\omega) \right] \, |\vec{\phi}_{\bq}|^2   \,.
\end{equation}
First, focus on the $\omega=0$ limit in order to understand the stability of the phase. In this regime, we derive a Landau--Ginzburg theory of the spin order parameter $\varphi_{\bq} = \vec{\phi}_{\bq}(\omega=0)$ in the spin-disordered phase.
We define the zero-frequency limit of the self energy $\Pi^0_{\bq}=\chi^0(\bq,0)$. This is purely real, and represents a shift of the stiffness due to RPA corrections from the positive mean-field value.
We can write the effective potential of the spin order parameter as
\begin{equation}\label{eq:app_vboson_sma}
   V(\varphi_{\bq}) \approx r_{\bq}\varphi_{\bq}^2  \equiv [m_0^{\, 2}(\bq) - 2\Pi^0_{\bq} ] \varphi_{\bq}^2 \, .
\end{equation}
When this stiffness $r_{\bq}$ becomes negative, there is an instability caused by the condensation of the parameter $\varphi_{\bq} > 0$. This coincides with the RPA pole condition at $\omega=0$.
We are hence able to predict the stability of the QSL to a competing magnetic ordered state (with ordering wavevector $\bq$) simply by calculating the RPA correction to its mean-field stiffness. Stability of this potential is equivalent to the real-part of the $\omega=0$ RPA susceptibility being positive.
Focusing on the effective potential for the spin order parameters with wavevectors $\bm{K}$ and $\bm{M}$ at low energy, we calculate $\Pi^0_{\bm{K}} = 0.389 \, J_1^{-1}$ and $\Pi^0_{\bm{M}} = 0.425 \, J_1^{-1}$.
The phase transitions occur when $r_{\bq} = 0$, and so we derive the transition points to be
\begin{align}
  J_{2,c}^{120}/J_1 &= \frac{1}{2}\left(1-\frac{1}{3\Pi^0_{\bm{K}}J_1}\right)
	= 0.0716 \, ,
	\label{eq:rc120}
  \\
	J_{2,c}^{{\mathrm{stripe}}}/J_1 &= \frac{1}{2\Pi^0_{\bm{M}}J_1} - 1
	= 0.1758 \, .
	\label{eq:rcstripe}
\end{align}
Returning to the $\omega>0$ problem, we will now calculate the mass gap of the spin fluctuations in the stable regime.
From the Hamiltonian~\eqref{eq:app_hsbosonenergy}, the Green's function of this boson is
\begin{widetext}
\begin{equation}
    G_\varphi(\bq,\omega) 
\,\,=\,\,
\begin{fmffile}{hs-3}
\begin{gathered}
\begin{fmfgraph*}(45,30)
    \fmfleft{i1}
    \fmfright{o1}
    \fmf{double}{i1,o1}
\end{fmfgraph*}
\end{gathered}
\end{fmffile}
\,\,=\,\, \frac{1}{m_0^2(\bq) -2\chi^0(\bq,\omega)}\,\,
=\,\,
\begin{fmffile}{hs-5}
\begin{gathered}
\begin{fmfgraph*}(45,30)
    \fmfleft{i1}
    \fmfright{o1}
    \fmf{dbl_dashes}{i1,o1}
\end{fmfgraph*}
\end{gathered}
\,\,+\,\,
\begin{gathered}
    \begin{fmfgraph*}(70,30)
    \fmfleft{i1}\fmfright{o1}
    \fmf{dbl_dashes,tension=2}{i1,w1}\fmf{dbl_dashes,tension=2}{w2,o1}
    \fmf{fermion,right=.5,tension=.5}{w1,w2}\fmf{fermion,right=.5,tension=.5}{w2,w1}
\end{fmfgraph*}
\end{gathered}
\end{fmffile}
\,\,+\,\,\cdots
\end{equation}\noindent
The quasiparticle manifests as a pole at $\omega=m_{\mathrm{RPA}}$ in the Greens function when $m_{0}^2(\bq) - 2\chi^0(\bq,m_{\mathrm{RPA}})=0$.
Expanding the susceptibility to second order around $\omega=0$, we find
\begin{align}
	\chi^0(\bq,\omega) &= \chi^0(\bq,0) + \omega^2 \left[\frac{1}{2}  \pdv[2]{\chi^0(\bq,0)}{\omega} \right] + \cdots
    \equiv
    \Pi^0_{\bq} + \omega^2 (\Pi_{\bq}^0)'' + \cdots \, .
\end{align}
Solving the pole condition perturbatively ${m}(\bq)\approx m_{\mathrm{RPA}}(\bq)$, we get the paramagnon Green's function is $G_\varphi(\bq,\omega) \approx \alpha^2/[{m}^2(\bq)-\omega^2]$ with approximate mass and normalization
\begin{equation}\label{eq:effmass_sma_new}
    	{m}^2(\bq) = \alpha^2_{\bq} \left [ m_0^{\, 2}(\bq) -2\Pi^0_{\bq} + \cdots\right]
        =\alpha^2_{\bq} r_{\bq} \, ,
\end{equation}
where $\alpha^2_{\bq} = 1/(2\Pi^0_{\bq})''$.
We numerically evaluate the constants $(\Pi^0_{\bm{K}})'' = 0.116 \, J_1^{-3}$ and $(\Pi^0_{\bm{M}})'' = 7.48 \, J_1^{-3}$, which are used to plot the gap in Fig.~\ref{fig1}(e).
Finally, we can derive the critical boson theory at the condensation transition to magnetic order at $\bm{K}$. To do this, one expands the susceptibility around $\bq=\bm{K}$, about which we observe it has a minimum
\begin{align}
	\chi^0(\bq,\omega) \equiv
    \Pi^0_{\bm{K}} + \frac{\alpha^{-2}_{\bm{K}}}{2} [\omega^2 - c^2_{\bm{K}}(\bq-\bm{K})^2] + \cdots \, ,
\end{align}
for an appropriate definition of $c_{\bm{K}}$.
Hence, we transform Eq.~\eqref{eq:app_hsbosonenergy} into Eq.~\eqref{eq:critical_boson} in the Results section after a field rescaling.

%
%

\subsection{Coupling to gauge field}
Can we better understand how the gauge fluctuations interact with the spinon-exciton mode?
A simple understanding can come from noticing that the coupling renormalises the bound state propagator
\cite{ghaemi2006}
\begin{equation}
G_\varphi(\bq,\omega) 
\,\, = \,\,
\begin{fmffile}{hs-3}
\begin{gathered}
\begin{fmfgraph*}(50,30)
    \fmfleft{i1}
    \fmfright{o1}
    \fmf{double}{i1,o1}
\end{fmfgraph*}
\end{gathered}
\,\,=\,\,
\begin{gathered}
\begin{fmfgraph*}(45,30)
    \fmfleft{i1}
    \fmfright{o1}
    \fmf{dbl_dashes}{i1,o1}
\end{fmfgraph*}
\end{gathered}
\,\,+\,\,
\begin{gathered}
    \begin{fmfgraph*}(70,30)
    \fmfleft{i1}\fmfright{o1}
    \fmf{dbl_dashes,tension=2}{i1,w1}\fmf{dbl_dashes,tension=2}{w2,o1}
    \fmf{fermion,right=.5,tension=.5}{w1,w2}\fmf{fermion,right=.5,tension=.5}{w2,w1}
\end{fmfgraph*}
\end{gathered}
\,\,+\,\,
\begin{gathered}
    \begin{fmfgraph*}(70,30)
    \fmfleft{i1}\fmfright{o1}
    \fmf{dbl_dashes,tension=2}{i1,w1}
    \fmf{dbl_dashes,tension=2}{w2,o1}
    \fmf{boson}{w1,w2}
\end{fmfgraph*}
\end{gathered}
\end{fmffile}
\,\,+\,\,\cdots
\end{equation}
This is exactly the same problem as studied in \cite{lee2019,seifert2024}. The frequency-dependent propagator leads to a hybridization and quasiparticle continuum in the previously sharp mode.
The Green's function of the paramagnon reads 
\begin{equation}
    G_\phi(\bq,\omega) = \frac{1}{m^2(\bq) - g_{\mathcal{S}}^2\langle \mathcal{S}_\alpha^\dagger \mathcal{S}\rangle(\bq,\omega)- \omega^2} \, .
\end{equation}
We can infer that the dispersive paramagnon mode $m^2(\bq)$ gets its energy renormalised by the gauge fluctuations, and it gets broadened.
The real and imaginary parts are respectively responsible for these phenomena.
The corrected energy and inverse lifetime are given as follows, where we evaluate the monopole susceptibility at the uncorrected mass:
\begin{align}
       \tilde{m}^2(\bm{K})
    &= {m}^2(\bm{K}) -g^2_{\mathcal{S}} c \, \cos[\pi(3/2-\Delta_\Phi)] \,|{m}(\bm{K}) |^{-(3-2\Delta_\Phi)}\, , \\
    \tau^{-1}(\bm{K})
    &= [g^2_{\mathcal{S}}  c / m(\bm{K})] \,
    \sin[\pi(3/2-\Delta_\Phi)]\, |m(\bm{K}) |^{-(3-2\Delta_\Phi)}
\end{align}
For $1<\Delta_\Phi <3/2$, the mode is drawn downwards and broadened \cite{seifert2024}.
We confirm this behaviour in our fermion model by introducing a fermion-monopole coupling to our RPA scheme in Supplementary Note 3.
\end{widetext}

\subsection{Nature of transition}
Here, we will argue the nature of the transition as the critical boson condenses when coupled to the monopoles. As discussed in the main text, this generates an effective monopole-monopole coupling Eq.~\eqref{eq:sphiphi}, which leads to a divergent interaction as the boson mass is taken to zero $m\to0$.
%
%
In this limit, the operators with zero charge can be constructed by considering the operator product expansion (OPE) of the monopole-antimonopole $\Phi^\dagger_{2\pi}\Phi^\dagga_{2\pi}$ \cite{zhang2025}.
The resulting operators transform in the following irreducible representations of $\mathrm{SO}(6)$:
\begin{equation}
    \bm{6} \,\otimes\, \bm{6} = \bm{1} \,\oplus\,\bm{15} \,\oplus\, \bm{20}' \, ,
\end{equation}
and have charge $q=0$ under $\mathrm{U}(1)_{\mathrm{topo}}$.
Here, $\bm{1}$ is the trivial SO(6) representation; the leading charge-zero operator here is the chiral fermion mass $M_{00}$. 
It is generated from the SO(6) trace-component $\sum_a\Phi^\dagger_a\Phi^\dagga_a$.
This transforms trivially under all ultraviolet  symmetries, and must be assumed to be irrelevant to avoid spontaneous chiral symmetry breaking and a stable conformal fixed point.
Next, the representation $\bm{15}$ is the SO(6)-adjoint, containing antisymmetric fermion bilinears $M_{\mu\nu}$.
This would be generated by an antisymmetric combination of monopoles.
Finally, $\bm{20}'$ contains the traceless symmetric combinations of monopoles.
Higher-rank representations coming from $\Phi^\dagger_{4\pi}\Phi^\dagga_{4\pi}$ and higher fluxes contribute at higher orders as well.
Further details and explicit forms of these operators in terms of monopoles given in Ref.~\cite{zhang2025}.
%

As we introduced above, the OPE of two monopoles $\Phi^\dagger\Phi$ produces charge-zero operators.
The symmetric combination in \eqref{eq:sphiphi} specifically leads to only four-fermion interactions appearing, from the sectors $\mathbf{1}\oplus\mathbf{20}'$.
The most relevant OPE channel from the above effective monopole interaction is the four-fermion interaction in the irrep $\bm{20}'$ of SU(4), which we now focus on.
The spin-triplet monopoles $\mathcal{S}_\alpha$ transform as $(\mathbf{1},\mathbf{3})$ under $\mathrm{SO}(3)_{\mathrm{valley}}\times \mathrm{SO}(3)_{\mathrm{spin}}$.
Hence, the product $\mathcal{S}_\alpha \times \mathcal{S}_\beta$ lives in the irrep $(\mathbf{1},\mathbf{1}) \oplus (\mathbf{1},\mathbf{3})\oplus (\mathbf{1},\mathbf{5})$, where $\mathbf{3}$ and $\mathbf{5}$ are the antisymmetric / symmetric-traceless irreps of $\mathrm{SO}(3)_{\mathrm{spin}}$.
The operator Eq.~\eqref{eq:sphiphi} takes the trace over the SO(3) index $\alpha=1\dots3$, and hence the resulting OPE produces an operator which transforms as $(\mathbf{1},\mathbf{1})$ under $\mathrm{SO}(3)_{\mathrm{valley}}\times \mathrm{SO}(3)_{\mathrm{spin}}$.
In terms of fermion operators, we can write this as follows
\begin{equation}\label{eq:chgnop}
   S_{\Phi\Phi} \sim g_{\mathcal{S}}^2 / m^2 \int\dd[3]{x} \left[(\overline{\psi}_i\vec{\sigma}\psi_i)^2 
   - (\overline{\psi}_i\vec{\tau}\psi_i)^2 \right]
   .
\end{equation}
This interaction is believed to be irrelevant, but will cause a phase transition as the spinon exciton condenses as $m^2\to 0$ which leads to the condensation of the order parameter $\overline{\psi}_i\vec{\sigma}\psi_i$.
This condensation leads to the proliferation of 120-degree magnetic order \cite{song2019}.
The interaction \eqref{eq:chgnop} is an O(3) Gross--Neveu QED$_3$ interaction term.
We predict it will lead to the same fixed-point as the chiral Heisenberg Gross--Neveu--Yukawa model \cite{dupuis2019} on the kagome lattice.
We therefore suggest that the spin liquid--120-degree order transition at $J_{2,c}/J_1 \approx 0.07$ on the triangular lattice is in this universality class.
This should be confirmed with a renormalization group study of the critical boson-monopole theory.

\section{XXZ Anisotropy}\label{app3}
We take the XXZ Hamiltonian Eq.~\eqref{eq:xxz} and focus on the XY term:
\begin{multline}
    \mathcal{H}_{\mathrm{XY}} = \sum_{\langle\bm{i}\bm{j}\rangle} J_{\bm{i}\bm{j}}\, \left( S^x_{\bi}S^x_{\bj} + S^y_{\bi}S^y_{\bj}  \right)  \\
    = \frac{1}{2} \sum_{\langle\bm{i}\bm{j}\rangle} J_{\bm{i}\bm{j}}\, \left( S^+_{\bi}S^-_{\bj} + S^-_{\bi}S^+_{\bj}  \right) 
    \, .
\end{multline}
Now using $S^+_{\bi} = f^{\dagger}_{\bi\uparrow}f^{\dagga}_{\bi\downarrow}$, we can write
\begin{equation}
    \mathcal{H}_{\mathrm{XY}} 
    = \frac{1}{2} \sum_{\langle\bm{i}\bm{j}\rangle} J_{\bm{i}\bm{j}}\, \left( 
    f^{\dagger}_{\bi\uparrow}f^{\dagga}_{\bj\uparrow} f^{\dagger}_{\bj\downarrow}f^{\dagga}_{\bi\downarrow}
    + 
        f^{\dagger}_{\bi\downarrow}f^{\dagga}_{\bj\downarrow} f^{\dagger}_{\bj\uparrow}f^{\dagga}_{\bi\uparrow}
    \right) 
    \, .
\end{equation}
Now we can mean-field decouple these as before, using $\chi_{\bij} = \langle f^\dagger_{\bm{i}\alpha}f_{\bm{j}\alpha}^\dagga \rangle = 2\langle f^\dagger_{\bm{i}\uparrow}f_{\bm{j}\uparrow}^\dagga \rangle$, to find
\begin{equation}
{\mathcal{H}}_{\rm 0} 
=-\frac{J_1}{4}\sum_{ \bm{i}\bm{j}} (\chi_{\bm{i}\bm{j}} f^\dagger_{\bm{j}\alpha} f^\dagga_{\bm{i}\alpha} + {\rm h.c.} ) \, .
\end{equation}
This is the same $\pi$-flux Ansatz as before, except with half the overall prefactor. Combining this result with the Heisenberg case, some simple algebra shows that the mean-field hopping of the general XXZ Hamiltonian Eq.~\eqref{eq:xxz} is then
\begin{equation}
    t_{\bm{i}\bm{j}} = \frac{1}{4} (1+\Delta) J_{\bij} \chi_{\bij} \, .
\end{equation}
This changes the normalization of the parton bands, and hence changes the range of stability of the QSL phase.
Note that we use the same Gutzwiller renormalization in the XXZ model; this approximation potentially breaks down away from $\Delta=1$, adjusting the resulting structure factor predictions.

The second effect of XXZ anisotropy is that the interaction matrix becomes spin-component dependent. For this reason, we calculate the general spin susceptibility matrix
\begin{equation}\label{eq:suscgenij}
    \chi^{ij}(\bm{q},\iu\omega)
    = \frac{1}{N} \int \dd{\tau} \, \eu^{\iu\omega \tau} \langle T_\tau \,  {S}_{\bm{q}}^i(\tau) {S}_{-\bm{q}}^j(0)\rangle\, ,
\end{equation}
which contains both the diagonal element $\chi^{zz}(\bm{q},\iu\omega)$ component, as well as transverse component $\chi^{\pm}(\bm{q},\iu\omega)= \chi^{xx}(\bm{q},\iu\omega) + \chi^{yy}(\bm{q},\iu\omega) + \iu [ \chi^{xy}(\bm{q},\iu\omega)- \chi^{yx}(\bm{q},\iu\omega)]$.
The RPA equation is also modified with a spin-direction dependent interaction which distinguishes the components, $J_{\bq} ( S^x_{\bq} S^x_{-\bq} + S^y_{\bq} S^y_{-\bq} + \Delta S^z_{\bq} S^z_{-\bq})$.
In this basis, the RPA interaction becomes 
\begin{equation}
    U_{\bq}^{ij} = 
    -2J_{\bq} \operatorname{diag}(1, 1, \Delta) \, .
\end{equation}
This split interaction is directly responsible for the different energy of the paramagnon in different channels.

The universality class of the transition is modified by the XXZ anisotropy.
The three components of the paramagnon have different energies, and therefore only one or two components condense simultaneously. This leads to a restriction over the SO(3) index in the induced monopole-monopole interaction term Eq.~\eqref{eq:sphiphi}.
We take only $\sum_{\alpha=1,2} \mathcal{S}^\dagger_\alpha \mathcal{S}^\dagga_\alpha$ for the case of easy-plane anisotropy, and $\mathcal{S}^\dagger_3 \mathcal{S}^\dagga_3$ for easy-axis anisotropy.
Breaking $\mathrm{SO}(3)_{\mathrm{spin}}$ symmetry means that the OPE also mixes in operators in the symmetric-traceless irrep $(\mathbf{1},\mathbf{5})$.
The result is the term $(\overline{\psi}_i\vec{\sigma}\psi_i)^2$ in Eq.~\eqref{eq:chgnop} is replaced by the terms \diff{$(\overline{\psi}_i\sigma_3\psi_i)^2$ and $(\overline{\psi}_i\sigma_1\psi_i)^2 + (\overline{\psi}_i\sigma_2\psi_i)^2$} with reduced-symmetries, respectively.
These correspond to transitions with a remaining \diff{$\mathbb{Z}_2$ or U(1)} symmetries \cite{dupuis2022} \diff{in the chiral-Ising and chiral-XY universality classes.}
\diff{This identification is confirmed by exact diagonalisation of the vector-chirality correlations on small clusters~\cite{iqbal2026a}.}

\section{Chiral QSL}\label{app4}

Next, we demonstrate the generality of the self-consistent approach by perturbing the U(1) Dirac QSL with a scalar spin chirality (SSC) term Eq.~\eqref{eq:ssc}.
The SSC term induces six-fermion interactions. We can split the interaction to find \cite{maity2025}
\begin{equation}
  H_\chi = J_\chi \vec{S}_{\bi}\cdot (\vec{S}_{\bj}\cross \vec{S}_{\bk})
  = H_{\bi\bj\bk} + H_{\bj\bk\bi} + H_{\bk\bi\bj} \, , 
\end{equation}
where the explicit interaction in the parton basis is given by 
\begin{align}
  H_{\bi\bj\bk} &= \frac{\iu J_\chi}{2} \left( S^+_{\bi}S^-_{\bj} - S^-_{\bi}S^+_{\bj} \right) S^z_{\bk} \\ 
  \nonumber
  &= \frac{\iu J_\chi}{4} \Big[ 
	  f^\dagger_{\bi\uparrow} f^\dagga_{\bk\uparrow}f^\dagger_{\bk\uparrow} f^\dagga_{\bj\uparrow} f^\dagger_{\bj\downarrow} f^\dagga_{\bi\downarrow}
\\ &\quad\quad
	- f^\dagger_{\bi\downarrow} f^\dagga_{\bj\downarrow}f^\dagger_{\bj\uparrow} f^\dagga_{\bk\uparrow} f^\dagger_{\bk\uparrow} f^\dagga_{\bi\uparrow}
	+ (\uparrow\leftrightarrow\downarrow)
   \Big] .
   \label{eq:Hinter_ssc}
\end{align}
As is familiar, we can now decouple the interacting Hamiltonian using parton mean-field theory.
The Hamiltonian takes the following form \cite{banerjee2023} 
\begin{multline}\label{eq:H0chiraldecomp}
  \mathcal{H}_0 = -\frac{1}{2} \sum_{\bi\bj}J_{\bij} [(\chi_{\bij} f^\dagger_{\bj\alpha} f^\dagga_{\bi\alpha} + \hc) - |\chi_{\bij}|^2] \\
  + \frac{3\iu }{4} \sum_{\bi\bj\bk} J_\chi (
  \chi_{\bi\bk}\chi_{\bk\bj} f^\dagger_{\bj\alpha} f^\dagga_{\bi\alpha} + 
  \chi_{\bi\bk}\chi_{\bj\bi} f^\dagger_{\bk\alpha} f^\dagga_{\bj\alpha}
  \\
  +\chi_{\bk\bj}\chi_{\bj\bi} f^\dagger_{\bi\alpha} f^\dagga_{\bk\alpha}
  - \chi_{\bi\bk}\chi_{\bk\bj}\chi_{\bj\bi}
   - \hc) \, .
\end{multline}
In order to work with this mean-field Hamiltonian, we assume that the $\chi_{\bij}$ form a $[\pi,0]$ flux Ansatz.
The mean-field parameters $\chi_{\bij}$ in this context are all real, shown in Fig.~\ref{fig1}(b).
The mean-field hoppings that they induce in Eq.~\eqref{eq:H0chiraldecomp} will become complex. 
We show in Supplementary Note 4 that this modifies the $[\pi,0]$ Ansatz to have hopping amplitude $t$ with staggered fluxes $[\pi-\theta,\theta]$, where
\begin{align}
  t = \frac{J_1\chi}{2} \sqrt{1+\left( \frac{3\chi J_\chi}{2J_1} \right)^2}\, , \\
   \theta= -3\arctan(\frac{3\chi J_\chi}{2J_1}) \, .
\end{align}
Ref.~\cite{hu2016variational} used a variational method to calculate the optimal phase $\theta$ as a function of $J_\chi/J_1$ up to $J_\chi/J_1 = 0.3$ where data are available. 
Our method reproduces the numerical results for the hopping phase $\theta$ across this range; as a result of this success we extend our method up to $J_\chi/J_1 = 0.5$.

We will evaluate the \emph{interacting} susceptibility of the theory in the presence of SSC interactions. We calculate the susceptibility in the spin channel using the random phase approximation.
The RPA sums over a series of bubble diagrams, where the interactions are four-fermion vertices. 
The SSC interaction takes the form
\begin{equation}
  H_\chi
= 
  J_\chi  \,  \frac{1}{8}\epsilon_{ijk} \sigma^i_{\alpha\beta}  \sigma^j_{\gamma\delta}\sigma^k_{\sigma\rho}  \sum_{\bi\bj\bk} 
f^\dagger_{\bi\alpha} f^\dagga_{\bi\beta} \, 
f^\dagger_{\bj\gamma} f^\dagga_{\bj\delta}\,
f^\dagger_{\bk\sigma} f^\dagga_{\bk\rho} \, .
    \label{eqL:interaction_xybasis_ssc}
\end{equation}
The contribution of this interaction at first order would give an antisymmetric, off-diagonal (in spin indices $ij$) contribution the susceptibility, since $\langle f^\dagger_{\bi\alpha} \sigma_{\alpha\beta}^\dagga f^\dagga_{\bi\beta} \rangle=0$ it must vanish in the mean-field.
We show in Supplementary Note 4 that the leading contribution comes in at second order, which
leads to an effective spin-spin interaction
\begin{equation}
    H_{\mathrm{eff}} = -\frac{1}{2} \int \dd{\tau} H_\chi(0)H_\chi(\tau) 
    \approx 
    \frac{J_\chi^2}{8J_1} \sum_{\bij}\vec{S}_{\bi}\cdot\vec{S}_{\bj} \, ,
\end{equation}
with $\bij$ along both nearest and next-nearest-neighbor bonds.
We comment that the effect of this interaction is minor compared to the stabilising effect of introducing a fermion gap. Overall, the $J_\chi$ interaction tends to stabilise the chiral QSL phase compared to the Dirac case.

\section*{Data availability} 
Data and code are available from the authors upon reasonable request.

\bibliographystyle{unsrt}
\bibliography{main}


\end{document}





\title{\vspace{-2.0cm}\Large\textbf{Supplementary Information:}\\{``Dynamics and stability of U(1) spin liquids beyond mean-field theory:
Triangular-lattice $J_1$--$J_2$ Heisenberg model''}}
	
\author[1,2,3]{Josef Willsher}
\author[1,2,4]{Johannes Knolle}

\affil[1]{Technical University of Munich, TUM School of Natural Sciences, Physics Department, 85748 Garching, Germany}
\affil[2]{Munich Center for Quantum Science and Technology (MCQST), Schellingstr. 4, 80799 M{\"u}nchen, Germany}
\affil[3]{Max-Planck-Institut f\"{u}r Physik komplexer Systeme, 01187 Dresden, Germany}
\affil[4]{Blackett Laboratory, Imperial College London, London SW7 2AZ, United Kingdom}

\date{}

\maketitle

\section*{Supplementary Note 1: Random phase approximation for QSLs}

\subsection*{Hamiltonian and interactions}





We are interested in Heisenberg spin models, where in the Abrikosov fermion representation the model contains pure four-fermion interaction terms. In real space, this is 
\begin{multline}
    {\cal H} = \sum_{\langle\bm{i}\bm{j}\rangle} J_{\bm{i}\bm{j}}\, \vec{S}_{\bm{i}} \cdot \vec{S}_{\bm{j}} 
    = \sum_{\langle\bm{i}\bm{j}\rangle} J_{\bm{i}\bm{j}}\left(\frac{1}{2}f^\dagger_{\bi\alpha}\vec{\sigma}_{\alpha\beta}f^\dagga_{\bi\beta}\right) \cdot \left(\frac{1}{2}f^\dagger_{\bj\alpha'}\vec{\sigma}_{\alpha'\beta'}f^\dagga_{\bj\beta'}\right)
    = \frac{1}{2} \sum_{\langle\bm{i}\bm{j}\rangle} 
    J_{\bm{i}\bm{j}} \left(f_{\bm{i}\alpha}^\dagger  f_{\bm{i}\beta}^\dagga f_{\bm{j}\beta}^\dagger f^\dagga_{\bm{j}\alpha} - \frac{1}{2} f_{\bm{i}\alpha}^\dagger f_{\bm{i}\alpha}^\dagga f_{\bm{j}\beta}^\dagger f_{\bm{j}\beta}^\dagga \right)
    \, .
    \label{eq:interacting_H}
\end{multline}
In momentum space the model has no fermion bilinear terms.

The $\pi$-flux Ansatz and its two-site unit cell is drawn in Supplementary Fig.~\ref{fig:piflux-fig}\textbf{a}. The fermion hopping in real space is chosen to be $\pm t$ on the black/blue bonds.
We construct our theory in terms of sublattice fermions $f_{\bi,X}$ where $X=A,B$. The lattice unit vectors are $\bm{a}_1 = (2,0)$, $\bm{a}_2 = (1/2, \sqrt{3}/2)$. Consequently, the reciprocal space is shown in Supplementary Fig.~\ref{fig:piflux-fig}\textbf{b}; the inverse lattice unit vectors are $\bm{Q}_1 = \pi/a(1,-1/\sqrt{3})$ and $\bm{Q}_2= (0,4\pi/a\sqrt{3})$.
We henceforth use a basis of momenta $\bm{k} = k_1\bm{Q}_1 + k_2\bm{Q}_2$ and the reduced Brillouin zone is drawn as the gray dashed rectangle.

\begin{figure}[b]
\centering
\includegraphics{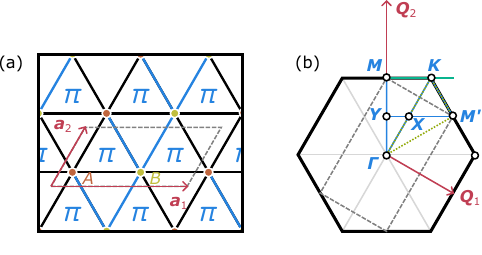}
\caption{(a) $\pi$-flux Ansatz with $A,B$ sites within the unit cell highlighted. Lattice vectors $\bm{a}_{1,2}$ and (b) reciprocal lattice unit vectors $\bm{Q}_{1,2}$ \label{fig:piflux-fig}}
\end{figure}

The Hamiltonian (which is diagonal in spin) can be written in terms of a spinor $\Psi_{\bm{k},m}$
\begin{equation}\label{eq:hammomspace}
    \mathcal{H}_0 = -\sum_{\bm{k}}
    \Psi^\dagger_{\bm{k}}
    H(\bm{k})
    \Psi_{\bm{k}}^\dagga
\end{equation}
with combined spin-sublattice index $m = (\alpha,X)$.
The spinor components are explicitly $\Psi_{\bm{k}} = (f_{\bm{k}\uparrow,A}, f_{\bm{k}\downarrow,A},f_{\bm{k}\uparrow,B}, f_{\bm{k}\downarrow,B})^T$ and
the Bloch Hamiltonian in this basis is
\begin{equation}
    H_{\alpha\beta}^{XY}(\bm{k})= t\,\delta_{\alpha\beta}\,M_{\bm{k}}^{XY}
\end{equation}
where explicitly the phases are given by
\begin{align}
    M^{AA}_{\bm{k}}&= -M^{BB}_{\bm{k}} = 2 \cos(k_2) \nonumber\\
    M^{AB}_{\bm{k}}&= 1 + \eu^{-\iu k_1} + \eu^{\iu (k_2-k_1)} - \eu^{-\iu k_2}\, ,
    \label{eq:subl_blochham}
\end{align}
and where $M^{XY}_{\bm{k}} =  (M^{YX}_{\bm{k}})^*$.
The eigenvalues of this matrix $H_{mn}(\bm{k})$ are written $\varepsilon^m_{\bm{k}}$ with corresponding eigenvectors $\alpha^m_{\bm{k}}$.
The overall normalisation of this mean-field Hamiltonian is fixed self-consistently by evaluating $t = \frac{J_1}{2N} \sum_{\bm{k}\alpha}
    \langle f_{\bm{k}\alpha, A}^\dagger f_{\bm{k}\alpha, B}^\dagga\rangle$.

We can write this mean-field theory in a self-consistent diagrammatic approach.
The free propagator of this model is therefore the trivial massless fermion 
\begin{equation}
        \mathcal{G}_0(\bm{q},\tau)^{XY}_{\alpha\beta} = -\langle T_\tau\, f_{\alpha X, \bm{q}}(0) \,f^\dagger_{\beta Y, \bm{q}}(\tau) \rangle \, ,
\end{equation}
which in Fourier space gives
\begin{equation}
    \mathcal{G}_0(\bq,{\omega+\iu\eta})^{mn}
    = -\frac{\alpha^{m}(\alpha^{n})^*}{{\omega+\iu\eta}}
\end{equation}
From hereon, it is standard to evaluate different mean-field decoupling channels of the interactions. The resulting quadratic theories can be categorized based on their resulting gauge structure~\cite{wen2004quantum}.
%
Self-consistent mean fields can be expressed diagrammatically as the self-energy. 
%
We depict the four-fermion interaction using the following Feynman vertex, where the interaction is drawn as a line between points $\bm{i}$,$\bm{j}$ and the fermion legs are 
\begin{fmffile}{vertex}
\begin{equation}\label{eq:Hint_diagram}
\mathcal{H}_{\mathrm{I}}
\,\,=\,\,
\begin{gathered}
\begin{fmfgraph*}(80,30)
 \fmfleftn{l}{2}\fmfrightn{r}{2}
  \fmflabel{$\alpha$}{l2}\fmflabel{$\beta$}{l1}
  \fmflabel{$\gamma$}{r1}\fmflabel{$\delta$}{r2}
 \fmfdot{K1}\fmfdot{K2}
 \fmflabel{$\bm{i}$}{K1}
 \fmflabel{$\bm{j}$}{K2}
 \fmf{fermion}{l1,K1}\fmf{fermion}{K1,l2}
 \fmf{dashes,tension=1}{K1,K2}
 \fmf{fermion}{r2,K2}\fmf{fermion}{K2,r1}
\end{fmfgraph*}
\end{gathered}
\,\,=\,\,
\frac{1}{2}J_{\bm{i}\bm{j}} \, f_{\bm{i}\alpha}^\dagger f^\dagga_{\bm{i}\beta}f_{\bm{j}\gamma}^\dagger f_{\bm{j}\delta}^\dagga \left( \delta_{\alpha\delta}\delta_{\beta\gamma} - \frac{1}{2}\delta_{\alpha\beta}\delta_{\gamma\delta} \right)\, .
\end{equation}
\end{fmffile}
This interaction leads to a fermion self-energy $\Sigma(\bq,\omega)$, which we write in real-space. The Fock term contributes to the self-energy as follows
\begin{fmffile}{fockterms}
\begin{equation}\label{eq:selfcons_greensfn}
\Sigma_{\bi\bj}^{\alpha\delta}
= 
-\frac{1}{2} J_{\bm{i}\bm{j}}\delta^{\alpha\delta}  \langle f^\dagger_{\bm{i}\beta} f^\dagga_{\bm{j}\beta}\rangle
\,\,=\,\,
\alpha\!\!
\begin{gathered}
\begin{fmfgraph}(50,30)
\fmfleft{i}
\fmfright{o}
\fmf{phantom,tension=5}{i,v1}
\fmf{phantom,tension=5}{v2,o}
\fmf{dashes,left,tension=0.4}{v1,v2}
\fmf{fermion}{v1,v2}
\end{fmfgraph}
\end{gathered}
\!\!\delta
\end{equation}
\end{fmffile}
The Hartree terms do not contribute to the self energy, since $\langle f^\dagger_{\bi\alpha}\sigma^z _{\alpha\beta}f^\dagga_{\bi\beta}\rangle = 0$ (and we ignore a finite constant to the overall energy).
The fermion propagator here is the interacting $\mathcal{G}(\bq,\omega)$, understood to be defined self-consistently in terms of this self-energy.
Recalling that the `bare' free Hamiltonian is zero for Heisenberg interactions in the parton basis, we find the following effective quadratic theory
\begin{equation}
    \mathcal{H} = \sum_{\bi\bj} \left(-t_{\bi\bj}f^\dagger_{\bj\alpha}f^\dagga_{\bi\alpha} + \hc + \mathcal{H}_{\mathrm{I}} \right)\, .
\end{equation}
where the hopping is purely generated by the (spin-diagonal) fermion self energy $t_{\bij} = \Sigma_{\bij}^{\alpha\alpha} = J_{\bij}\langle f^\dagger_{\bm{i}\beta} f^\dagga_{\bm{j}\beta}\rangle / 2$. This agrees with the (non-diagrammatic) mean-field theory of Refs.~\cite{wen1991,wen2002}.


\subsection*{Structure factor}
To evaluate the dynamical structure factor, we first make use of the fluctuation-dissipation theorem \cite{auerbach1998,giamarchi2003}. This relates the spectral function (which measures fluctuations) to the imaginary part of the dynamical spin susceptibility (which measures dissipative response).
The theorem states, up to a proportionality constant, that
\begin{equation}\label{eq:app_strucfac}
     S(\bm{q},\omega) = \operatorname{Im}{\chi(\bq,\omega)}
\end{equation}
where the susceptibility is defined in imaginary time
\begin{equation}\label{eq:susc_general_togf}
    \chi(\bq,\iu \omega_n) = \frac{1}{N}\sum_{\bi\bj} \eu^{-\iu \bm{q}\cdot (\bm{r}_{\bi} - \bm{r}_{\bj})} \int \dd{\tau} \, \eu^{-\iu\omega_n \tau} S_{\bi\bj}(\tau)\,.
\end{equation}
The time-ordered dynamical spin correlation function is
\begin{align}
    S_{\bi\bj}(\tau) =\langle T_\tau  \, S_{\bi}^z(\tau)S_{\bj}^z(0)
    \rangle  \, .
\end{align}
After evaluating correlation functions, we must analytically continue the frequencies back to real time $\iu \omega \to \omega + \iu \eta$.

The basic quantity of interest is the interacting fermion susceptibility. To evaluate this, we move Eq.~\eqref{eq:susc_general_togf} into the sublattice $X=A,B$ basis
\begin{equation}
        [\chi(\bm{q},\omega)]^{XY}
    = \frac{1}{N} \int \dd{\tau} \, \eu^{-\iu\omega t} \langle T_\tau \,  {S}_{X, \bm{q}}^z(\tau) {S}_{Y, -\bm{q}}^z(0)\rangle\, .
\end{equation}
Next, we move to the Abrikosov fermion representation
\begin{align}
    S_{X,\bq}^z &= \frac{1}{2}\sum_{\bm{k}}f^\dagger_{\alpha X,\bm{k}}(\sigma^{z})^\dagga_{\alpha\beta}f^\dagga_{\beta X,\bm{k}+\bq} 
    \\
    &= \frac{1}{2}\sum_{\bm{k}}\Psi_{\bm{k}}^\dagger\cdot(\sigma^{z}_X)\cdot \Psi_{\bm{k}+\bq}^\dagga
    \, .
\end{align}
Here once again, we move into the combined index $m=(\alpha,X)$. The sublattice Pauli matrix $\sigma^{z}_X$ only has non-zero elements when $m,n$ are in sublattice $X$. In other words, $\sigma^{z}_X = \sigma^z\otimes P_X$, where $P_A = \diag(1,0)$ and $P_B = \diag(0,1)$ are the projectors onto the $A,B$ sublattices.

We find the following, after imposing momentum conservation
\begin{equation}\label{eq:susceptibilitydef_diag}
\begin{fmffile}{susceptibilitydef}
[\chi(\bm{q},\omega)]^{XY} =
%
\quad\quad\quad
\begin{gathered}
\begin{fmfgraph*}(60,40)
 \fmfleftn{l}{1}\fmfrightn{r}{1}
  \fmflabel{$\frac{1}{2}\sigma_X^z$}{l1}\fmflabel{$\frac{1}{2}\sigma_Y^z$}{r1}
\fmfrpolyn{shaded}{G}{4}
 \fmf{fermion,right=0.25,tension=2}{l1,G1}\fmf{fermion,right=0.25,tension=2}{G2,l1}
 \fmf{fermion,right=0.25,tension=2}{r1,G3}\fmf{fermion,right=0.25,tension=2}{G4,r1}
\end{fmfgraph*}
\end{gathered}
\quad\quad
=
       \frac{1}{4N}
       (\sigma^{z})_{\alpha\beta}(\sigma^{z})_{\gamma\delta} \int \dd{\tau} \, \eu^{-\iu\omega_n \tau} \sum_{\bm{k} } 
       \langle T_\tau [f^\dagger_{\alpha X, \bm{k}} \, f^\dagga_{\beta X,\bm{k}+\bm{q}}](\tau)  \, [
    f^\dagger_{\gamma Y, \bm{k}'} \, f^\dagga_{\delta Y, \bm{k}'-\bm{q}}](0)\rangle\, .
\end{fmffile}
\end{equation}
This susceptibility matrix still has sublattice indices $XY$, since it is defined in momentum space in terms of gauge-dependent fermions. This object itself should however be gauge invariant since it represents a physical observable.
Since this object is a Fourier transform of a time-ordered Green's function \eqref{eq:susc_general_togf}, we can get the total susceptibility by summing over sublattices, including a relative phase
\begin{equation}\label{eq:app_susc_sumphases}
    \chi(\bq,\omega) = \sum_{X,Y=A,B}\, \eu^{-\iu\bm{q}\cdot (\bm{r}_X - \bm{r}_Y)} [\chi(\bm{q},\omega)]^{XY}
    = \vec{R}_{\bq}^\dagger \cdot [\chi(\bm{q},\omega)] \cdot \vec{R}_{\bq}
\end{equation}
where 
\begin{equation}\label{eq:r_vec_def}
    \vec{R}_{\bq} = (1, \eu^{-\iu\bm{q}\cdot \bm{r}_B})^T 
    =  (1, \eu^{-\iu k_1/2})^T 
\end{equation}
where $\bm{r}_B$ is the offset of the $B$ site in the unit cell.

\subsubsection*{Bare susceptibility}

%

The zeroth order calculation we can perform to evaluate the susceptibility is the non-interacting approximation. This neglects all interactions which generated our mean-field Hamiltonian.
\begin{fmffile}{freebubble_approx}
\begin{equation}
[\chi^0(\bm{q},\omega)]^{XY} =
\quad\quad\quad
\begin{gathered}
\begin{fmfgraph*}(33,30)
 \fmfleftn{l}{1}\fmfrightn{r}{1}
  \fmflabel{$\frac{1}{2}\sigma_X^z$}{l1}\fmflabel{$\frac{1}{2}\sigma_Y^z$}{r1}
 \fmf{fermion,right=.5}{l1,r1}\fmf{fermion,right=.5}{r1,l1}
\end{fmfgraph*}
\end{gathered}
\label{eq:free_bubble_baredef}
\end{equation}
\end{fmffile}
Given that we are measuring longitudinal $\sigma^z$ spin fluctuations, one can think of the opposing lines in Eq.~\eqref{eq:free_bubble_baredef} as carrying equal spin $\uparrow,\downarrow$.
Defining each of the lines in terms of single-particle propagators
\begin{equation}
        \mathcal{G}^{XY}_{\alpha\beta}(\bm{q},\tau) = -\langle T_\tau\, f_{\alpha X, \bm{q}}(0) \,f^\dagger_{\beta Y, \bm{q}}(\tau) \rangle \, ,
\end{equation}
and after imposing $\bm{k}' = \bm{k}+\bm{q}$ the susceptibility reads
\begin{align}\label{eq:suscspin1}
        [\chi^0(\bm{q},\iu\omega_n)]^{XY}
    &= \frac{1}{4N}  \int \dd{\tau} \, \eu^{-\iu\omega_n \tau} 
    \sum_{\bm{k}} 
    (\sigma^{z})_{\alpha\beta} (\sigma^{z})_{\gamma\delta}\,
     \mathcal{G}^{YX}_{\delta\alpha}(\bm{k},\tau)
     \mathcal{G}^{XY}_{\beta\gamma}(\bm{k}+\bm{q},-\tau) \\
     &= \frac{1}{4N} \sum_{\bm{k},\iu\nu} \Tr[\sigma^z_X \cdot \mathcal{G}(\bm{k}+\bq) \cdot \sigma^z_Y \cdot \mathcal{G}(\bm{k})]
\end{align}
A Matsubara summation over the free fermion Greens functions and analytic continuation back into real-frequencies gives the free susceptibility
\begin{equation}
    [\chi^0(\bq,\omega+\iu\eta)]^{XY}
    =- \frac{1}{4 N} \sum_{\bm{k},mn}
    \frac{(n_{\bm{k}}^m -n_{\bm{k}+\bm{q}}^n) \,\mathcal{V}^{mn}_X (\mathcal{V}^{mn}_Y)^*}{\omega + \varepsilon_{\bm{k}}^m -\varepsilon_{\bm{k}+\bm{q}}^n +\iu \eta}
\end{equation}
where the Fermi--Dirac occupation functions are $n_{\bm{k}}^m = n_f(\varepsilon_{\bm{k}}^m)$ and the matrix elements defined as
\begin{equation}
    \mathcal{V}^{mn}_X
    = \vec{\alpha}_{\bm{k}}^\dagger \cdot(\sigma^z_X) \cdot\vec{\alpha}_{\bm{k}+\bq}
    =(\alpha^m_{\bm{k}})_{\alpha X}^* (\sigma^z)_{\alpha\beta} 
    (\alpha^n_{\bm{k}+\bm{q}})_{\beta X}.
    \label{eq:vmn_matels_trip}
\end{equation}

Note that we may also calculate generalized spin-spin susceptibilities by resolving the different spin components $i,j = x,y,z$, defined as time-ordered correlation functions of $S^i_{\bi}(\tau)S^j_{\bj}(0)$.
The resulting susceptibility matrix can then be written as a $6\times6$ matrix $[\chi(\bm{q},\omega)]^{ij}_{XY}$. The physical response function is then found by tracing over the diagonal spin indices.
This is often necessary when studying magnetically ordered states, such as stripe or 120-degree states on the triangular lattice \cite{willsher2023}. However this is not necessary here, since we consider a QSL Ansatz with SU(2) spin-rotation symmetry. This guarantees that the diagonal elements are equal, and the off-diagonal elements vanish $[\chi(\bm{q},\omega)]^{ij}_{XY} = \delta^{ij} \,[\chi(\bm{q},\omega)]^{XY}$. This was checked numerically, and will be important when considering interactions.

\subsubsection*{Interacting susceptibility}
Beyond the free-fermion approximation, the dynamical spin susceptibility will be corrected due to the Heisenberg interactions, written in Eq.~\eqref{eq:Hint_diagram}.
Here, we consider only the effect of interactions $[J(\bq)]_{XY}$ on dynamical response functions \cite{zhou2003,ghioldi2018dynamical,zhang2020}, given by the real-space nearest- and next-nearest-neighbor interaction $J_{\bi\bj}$.
Transforming the Heisenberg interaction into the sublattice Fourier basis, we express it as
\begin{equation}
    \sum_{\bq} \delta_{ij}J_{\bq}^{XY} \, S^{i}_{X,\bm{q}}S^{i}_{Y,-\bm{q}}
= 
     \frac{1}{4}\sigma^i_{\alpha\beta}  \sigma^j_{\gamma\delta} \sum_{\bm{k}\bm{k}',\bq} \delta_{ij}J_{\bq}^{XY}  \, 
    \left(f^\dagger_{\alpha X,\bm{k}+\bm{q}} f^\dagga_{\beta X,\bm{k}}\right)
   \left(f^\dagger_{\gamma Y,\bm{k}'-\bq} f^\dagga_{\delta Y,\bm{k}'}\right) \, .
    \label{eqL:interaction_xybasis_sigma}
\end{equation}
The Fourier interaction strength in momentum space is defined as
\begin{equation}
  J^{XY}_{\bq} = \sum_{\bm{\delta}_{\bij}:X\to Y} J_{\bij} \eu^{\iu \bq\cdot\bm{\delta}_{\bij}}
\end{equation}
where $\bm{\delta}_{\bij} = \bm{r}_{\bi}-\bm{r}_{\bj}$ is the vector connecting all interacting sites where interactions $J_{\bij} \neq 0$ are nonzero.
The matrix elements in the sublattice basis are explicitly given by
\begin{align}
    J^{AA}_{\bq} &= J^{BB}_{\bq} = 2J_1
     \cos(k_2) + 2J_2\cos(k_1-k_2) 
        \\
     J^{AB}_{\bq}
     &=
    J_1\left[1 + \eu^{-\iu k_1} + \eu^{\iu (k_2-k_1)}
     + \eu^{-\iu k_2} \right] \nonumber\\
     &+J_2 \left[\eu^{\iu k_2} + \eu^{\iu (2k_2 - k_1)}  + \eu^{-\iu(k_1+k_2)} + \eu^{-2\iu k_2}  \right]
\end{align}
and where $J^{AB}_{\bq} =  (J^{BA}_{\bq})^*$.
Note that the elements $J^{XY}$ differ from the Hamiltonian since the signs do not alternate on different bonds in the unit cell, and there is no self-consistent amplitude $t$, which is only non-zero on nearest-neighbor bonds in $M^{XY}$. 
Importantly, this means that next-nearest neighbor interactions are non-zero and will modify the physics of the model beyond the self-consistent mean-field approximation.

Evaluating the leading diagram in this series, we find
\begin{equation}
  [\chi(\bq,\omega)]^{XY} = [\chi^0(\bq,\omega)]^{XY} - 2  [\chi^0(\bq,\omega)]^{XX'} J^{X'Y'}_{\bq} [\chi^0(\bq,\omega)]^{Y'Y} + \cdots\, .
\end{equation}
We hence identify the interaction matrix
\begin{fmffile}{interactionnewbasis}
\begin{equation}\label{eq:app_interactionnewbasis}
U^{XY}_{\bq}
= 
\,\,
\begin{gathered}
\begin{fmfgraph*}(20,10)
 \fmfleftn{l}{1}\fmfrightn{r}{1}
 \fmfdot{l1}\fmfdot{r1}
 \fmf{dashes,tension=3}{l1,r1}
\end{fmfgraph*}
\end{gathered}\,\,
    =\,\, -2 J_{\bq}^{XY} \, ,
\end{equation}
\end{fmffile}\noindent
where $J^{XY}_{\bq}$ is the Fourier transform of the Heisenberg interaction $J_{\bi\bj}$ in the sublattice basis.
Finally, we define $U^{\mu\nu} = U^{XY}\delta_{ij}$ in terms of a combined spin-sublattice index. 
Beyond the leading correction, there is a whole series of bubble diagrams of the same form. This is expressed diagramatically as
\begin{fmffile}{susceptibilityseries}
\begin{equation}\label{eq:susceptibilityseries_diagrams}
\begin{gathered}
\begin{fmfgraph*}(60,30)
 \fmfleftn{l}{1}\fmfrightn{r}{1}\fmfrpolyn{shaded}{G}{4}
 \fmf{fermion,right=0.25,tension=2}{l1,G1}\fmf{fermion,right=0.25,tension=2}{G2,l1}
 \fmf{fermion,right=0.25,tension=2}{r1,G3}\fmf{fermion,right=0.25,tension=2}{G4,r1}
\end{fmfgraph*}
\end{gathered}
\,\,
=
\,\,
\begin{gathered}
\begin{fmfgraph*}(33,30)
 \fmfleftn{l}{1}\fmfrightn{r}{1}
 \fmf{fermion,right=.5}{l1,r1}\fmf{fermion,right=.5}{r1,l1}
\end{fmfgraph*}
\end{gathered}
\,\,
+
\,\,
\begin{gathered}
\begin{fmfgraph*}(80,30)
 \fmfleftn{l}{1}\fmfrightn{r}{1}
  \fmfdot{K1}\fmfdot{K2}
 \fmf{fermion,fermion,right=.5}{l1,K1}\fmf{fermion,fermion,right=.5}{K1,l1}
 \fmf{dashes,tension=4}{K1,K2}
 \fmf{fermion,fermion,right=.5}{r1,K2}\fmf{fermion,fermion,right=.5}{K2,r1}
\end{fmfgraph*}
\end{gathered}
\,\,
+
\,\,
\begin{gathered}
\begin{fmfgraph*}(125,30)
 \fmfleftn{l}{1}\fmfrightn{r}{1}
  \fmfdot{K1}\fmfdot{K2}
 \fmf{fermion,fermion,right=.5}{l1,K1}\fmf{fermion,fermion,right=.5}{K1,l1}
 \fmf{dashes,tension=4}{K1,K2}
   \fmfdot{K3}\fmfdot{K4}
 \fmf{fermion,fermion,right=.5}{K2,K3}\fmf{fermion,fermion,right=.5}{K3,K2}
 \fmf{dashes,tension=4}{K3,K4}
 \fmf{fermion,fermion,right=.5}{r1,K4}\fmf{fermion,fermion,right=.5}{K4,r1}
\end{fmfgraph*}
\end{gathered}
\,\,
+\cdots
\end{equation}
\end{fmffile}
%
In the random phase approximation, we will resum all terms of this form.
To make progress, we can re-express this series as a Dyson--Schwinger equation
\begin{fmffile}{susceptibilityresum}
\begin{equation}
%
\begin{gathered}
\begin{fmfgraph*}(60,30)
 \fmfleftn{l}{1}\fmfrightn{r}{1}
\fmfrpolyn{shaded}{G}{4}
 \fmf{fermion,right=0.25,tension=2}{l1,G1}\fmf{fermion,right=0.25,tension=2}{G2,l1}
 \fmf{fermion,right=0.25,tension=2}{r1,G3}\fmf{fermion,right=0.25,tension=2}{G4,r1}
\end{fmfgraph*}
\end{gathered}
\,\,
=
\,\,
\begin{gathered}
\begin{fmfgraph*}(33,30)
 \fmfleftn{l}{1}\fmfrightn{r}{1}
 \fmf{fermion,right=.5}{l1,r1}\fmf{fermion,right=.5}{r1,l1}
\end{fmfgraph*}
\end{gathered}
\,\,
+
\,\,
\begin{gathered}
\begin{fmfgraph*}(110,30)
 \fmfleftn{l}{1}\fmfrightn{r}{1}
 %
  \fmfdot{K1}\fmfdot{K2}
 \fmf{fermion,fermion,right=.5,tension=1.5}{r1,K1}\fmf{fermion,fermion,right=.5.,tension=1.5}{K1,r1}
 \fmf{dashes,tension=6}{K1,K2}
 %
%
 \fmfrpolyn{shaded}{G}{4}
 \fmf{fermion,right=0.25,tension=2}{K2,G1}\fmf{fermion,right=0.25,tension=2}{G2,K2}
 \fmf{fermion,right=0.25,tension=2}{l1,G3}\fmf{fermion,right=0.25,tension=2}{G4,l1}
\end{fmfgraph*}
\end{gathered}
\end{equation}
\end{fmffile}
or equivalently 
\begin{align}\label{eq:dyson_rpa}
    [\chi(\bq,\omega)]^{\mu\nu} &= [\chi^0(\bq,\omega)]^{\mu\nu} + [\chi(\bq,\omega)]^{\mu\rho} U_{\bq}^{\rho\sigma}[\chi^0(\bq,\omega)]^{\sigma\nu}  \\ 
     [\chi(\bq,\omega)]^{\mu\rho}\left ( \delta^{\rho\nu}  - U_{\bq}^{\rho\sigma}[\chi^0(\bq,\omega)]^{\sigma\nu}\right) &= [\chi^0(\bq,\omega)]^{\mu\nu} 
\end{align}
This can be solved to give the RPA susceptibility; using matrix notation we obtain
\begin{align}
     \chi(\bq,\omega) &= \chi^0(\bq,\omega)\cdot  \operatorname{inv}\left [ 1  - U_{\bq} \cdot \chi^0(\bq,\omega)\right] \\
     &= 
     \chi^0(\bq,\omega)\cdot  \operatorname{inv}\left [ 1  +2 J_{\bq} \cdot \chi^0(\bq,\omega)\right]
     \, .
\end{align}
The full susceptibility is plotted to produce the interacting responses shown in Fig.~3 of the main text.
%
%
The poles of the RPA matrix define the dispersion of the spinon-exciton bound state $m_{\mathrm{RPA}}(\bm{q})$ through
\begin{equation}
    \det\left[1- U_{\bq} \cdot {\chi}^0\big(\bq,m_{\mathrm{RPA}}(\bm{q})\big)\right] = 0 \, ,
\end{equation}
which are plotted as white dashed lines in Supplementary~Fig.~\ref{fig:appendixmodes}.
We plot as red lines the result of the single-mode approximation, Eq.~(40) in the main text.

\begin{figure}
\centering
\includegraphics[width=0.8\textwidth]{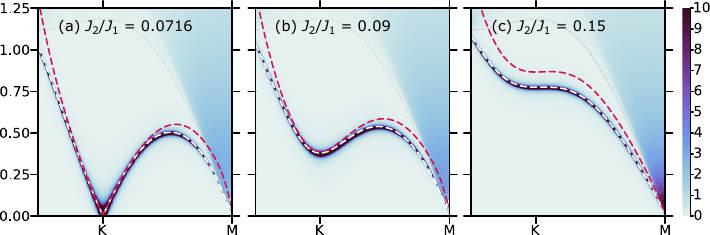}
\caption{Spectral function $S(\bq,\omega)$ for three different values of $J_2/J_1$ (a) $0.15$, (b) $0.09$, and (c) $0.0716$. White dotted line plots the zeroes of the RPA denominator $m_{\mathrm{RPA}}(\bm{q})$. Red dashed lines show the results of the single-mode approximation $m_{\varphi}(\bm{q})$.
}
\label{fig:appendixmodes}
\end{figure}

\subsubsection*{Real-space correlations}

Let us revisit the problem of spin-spin correlation functions in real space.
In terms of spin observables, it is
\begin{equation}\label{eq:gbij}
  g_{\bij} = \langle \vec{S}_{\bi} \cdot \vec{S}_{\bj}\rangle
\end{equation}
evaluated at equal time.
This is simply the dynamical spin-spin correlator $S_{\bij}(0)$ at equal time.
This, in turn, is the Fourier transform of the spin-spin susceptibility Eq.~\eqref{eq:susc_general_togf}.
Together, we can write
\begin{equation}
  g_{\bij} = \operatorname{Re}{\sum_{\bq} \eu^{\iu \bq\cdot(\bm{r}_{\bi}-\bm{r}_{\bi})}
  \int\dd{\omega}\chi(\bq,\omega)} \, .
\end{equation}
The real-space correlation function can be written as a Fourier transform over one momentum dimension of the susceptibility function, and a sum over the other momentum and frequency directions.
We choose a specific path of the correlation function along next-nearest-neighbor bonds $\bm{r}_{\bi}-\bm{r}_{\bi} = r(\bm{a}_1/2 + \bm{a}_2)$, see Supplementary Fig.~\ref{fig:piflux-fig}\textbf{a}.
This probes the excitations around the $\bm{K}$ point.
The resulting real-space correlations are shown in Supplementary Fig.~\ref{fig:realspace_rpa}\textbf{a} for a range of $J_2/J_1$. 
For large $J_2$, the response is approximately a power law with exponent $3.0(1)$.
For critical $J_2$, such that the spinon exciton gap $m$ goes to zero energy, the power law is $1.75(15)$.
For intermediate interaction strengths, there is a dominant exponential-like behaviour at short distances, as shown in the main text.

\begin{figure}
\centering
\includegraphics{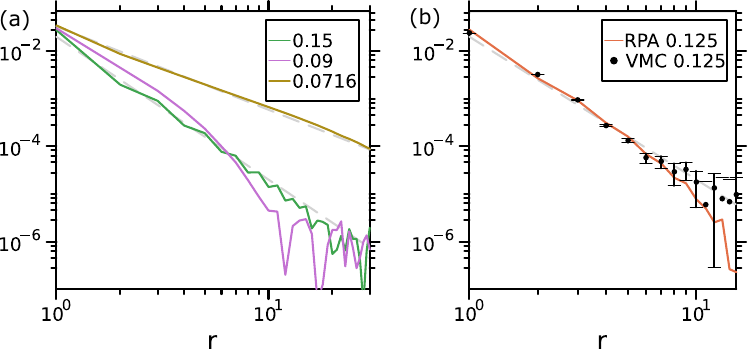}
\caption{Real-space correlation function $g(r)$ in the direction $\bm{a}_2+\bm{a}_1/2$. 
(a) RPA method for various $J_2/J_1=0.15$, $0.09$, and at the critical point $0.0716$.
(b) Comparison of the RPA method with VMC results of Ref.~\cite{ferrari2025} (c.f. Supplementary Figure 1) for $J_2/J_1=0.125$.
\label{fig:realspace_rpa}}
\end{figure}

We plot a direct comparison between VMC calculations \cite{ferrari2019} and our RPA method evaluated at $J_2/J_1=0.125$ in Supplementary Fig.~\ref{fig:realspace_rpa}\textbf{b}.
We are once again able to reproduce the numerical results very well. 
In fact, this result lets us put a bound on the contribution of monopoles to the static spin correlations.
If monopoles with scaling dimension $\Delta_\Phi\leq 1.25$ are present in spin correlation functions, they must have a prefactor that is of order $\mathcal{O}(10^{-3})$. Otherwise, they would dominate over the fermion contribution and our RPA result would clearly predict the wrong power law.
In the main text we consider a modified RPA scheme which includes a coupling to the monopoles (introced in Supplementary Note 3).

\newpage
\section*{Supplementary Note 2: Comparison to Numerical Methods}

\begin{figure}[h]
\centering
\includegraphics{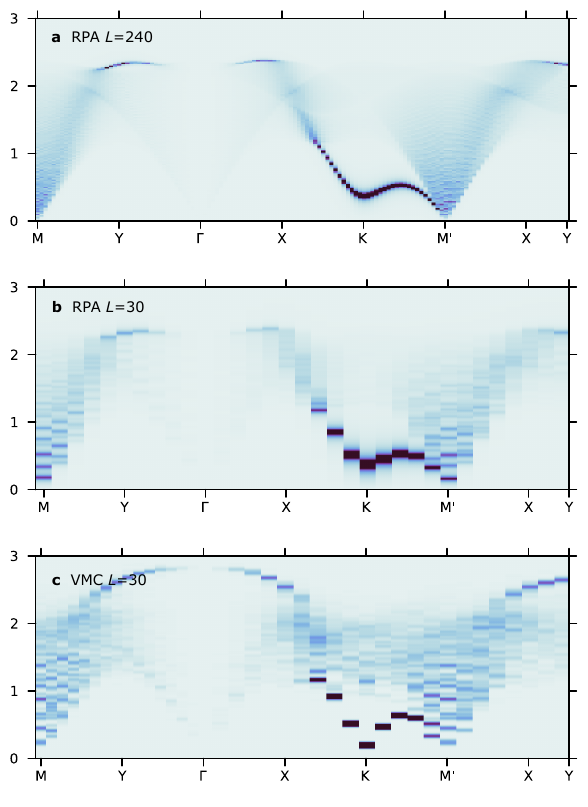}
\caption{\label{fig:comparison_L30}
\textbf{Comparison of the RPA method to VMC results from Ref.~\cite{ferrari2019}. Model parameters $J_2/J_1 = 0.09$.}
All energies are shown in units of $J_1$.
\textbf{a} RPA results along the direction $\bm{M}$-$\bm{\Gamma}$-$\bm{K}$-$\bm{M}$-$\bm{Y}$, performed on a $L=240$ system.
\textbf{b} RPA results along the same cut on a $L=30$ system; the Dirac cones have a finite-size gap.
\textbf{c} VMC results along the same cut, taken from Ref.~\cite{ferrari2019}.
}
\end{figure}

\begin{figure}[h]
\centering
\includegraphics[scale=0.8]{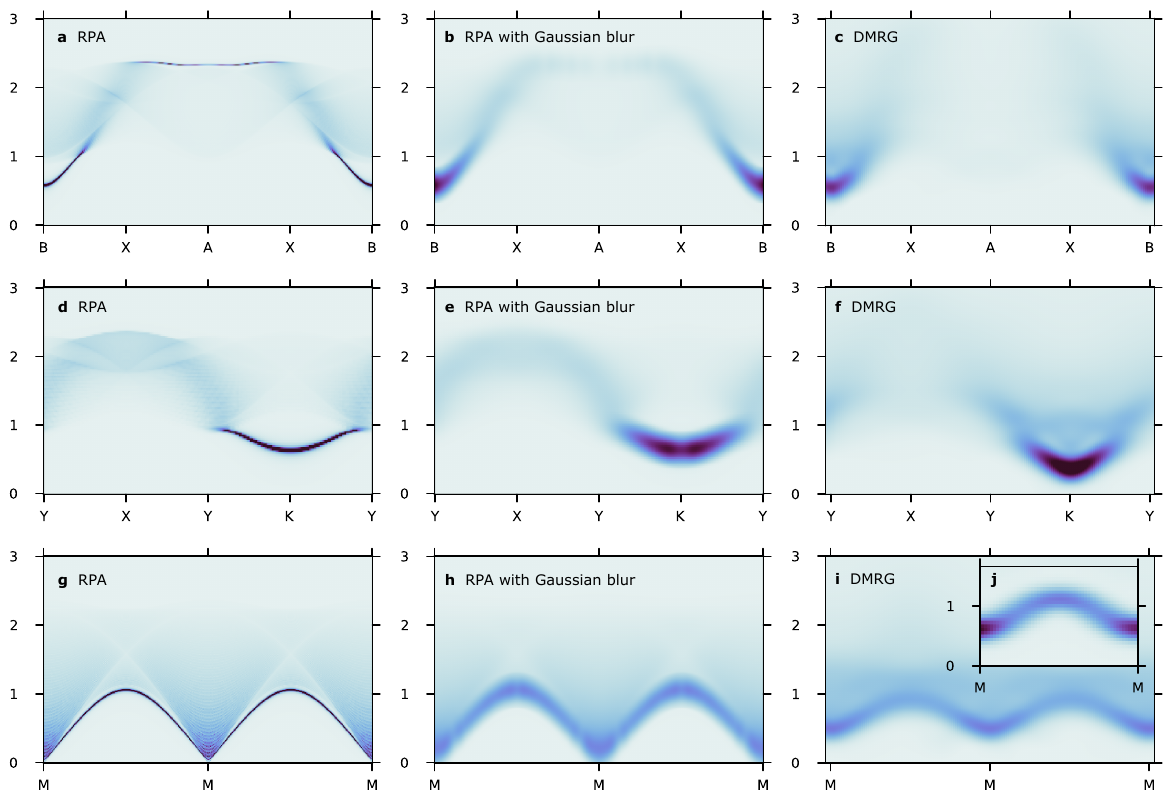}
\caption{\label{fig:dmrg1}
\textbf{Comparison of the RPA method to DMRG results from Ref.~\cite{drescher2025}. Model parameters $J_2/J_1 = 1/8$.}
RPA calculations performed on a $L=240$ system. All energies are shown in units of $J_1$.
\textbf{a} RPA results along the direction $\bm{X}$-$\bm{A}$-$\bm{X}'$.
\textbf{b} Blurred RPA results along the same cut, Gaussian broadening chosen to match the DMRG method.
\textbf{c} DMRG results along the same cut, taken from Ref.~\cite{drescher2025}.
Our results show that the low-energy peak of spectral intensity at $\textbf{B}$ is due to the spinon exciton leaving the two-spinon continuum.
\textbf{d-f} Results along the cut $\bm{X}$-$\bm{Y}$-$\bm{K}$.
Numerical results show a lower energy spinon exciton at the $\bm{K}$ point than RPA calculations, possibly suggesting the influence of gauge field fluctuations.
\textbf{g-i} Results along the cut $\bm{M}$-$\bm{M}'$.
The collective mode sits at low energies in the Dirac cone, which is gapped in $L_y=6$ DMRG calculations.
\textbf{j} Blurred RPA results on a $56 \times 6$ cylinder geometry along the $\bm{M}$-$\bm{M}'$ direction. We reproduce the finite-size gap and spectral weight distribution of DMRG.
}
\end{figure}
\clearpage

\begin{figure}[h]
\centering
\includegraphics[scale=0.8]{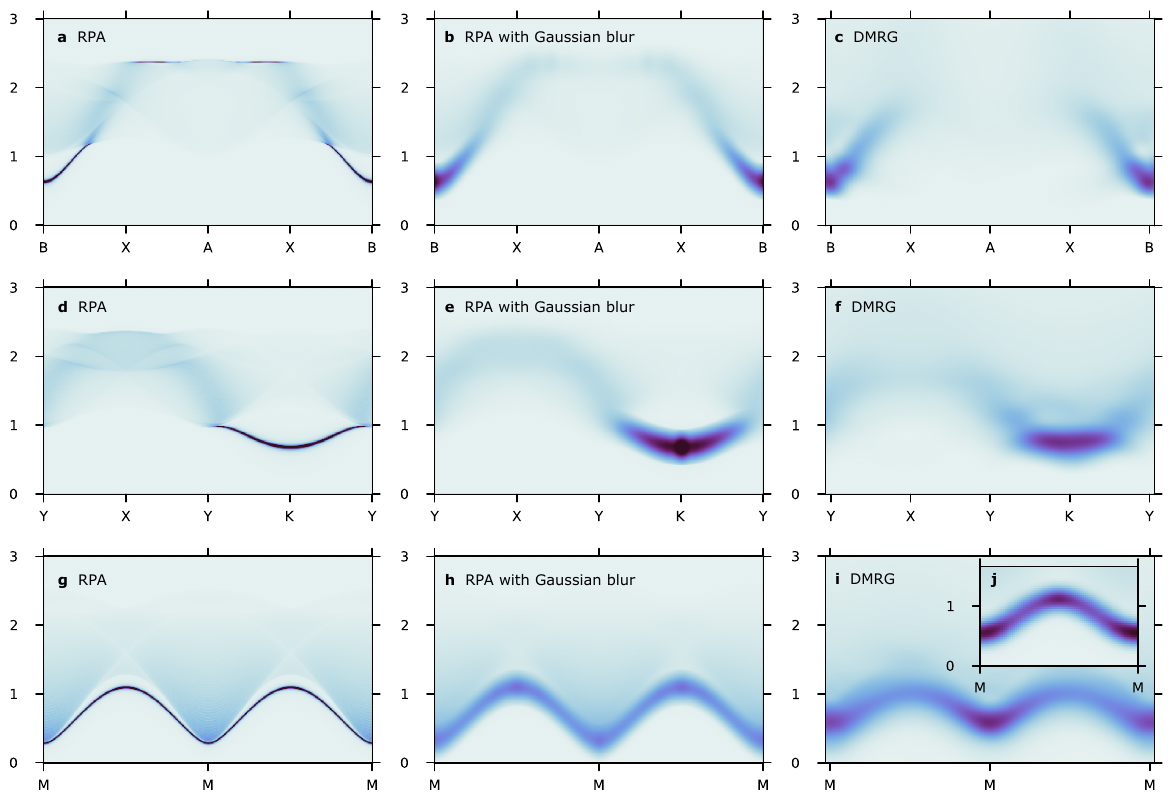}
\caption{\label{fig:dmrg2}
\textbf{Comparison of the RPA method to DMRG results from Ref.~\cite{drescher2025}.Model parameters $J_2/J_1 = 1/8$ and spin-scalar chirality $J_\chi/J_1=0.1$.}
RPA calculations performed on a $L=240$ system. All energies are shown in units of $J_1$.
\textbf{a} RPA results along the direction $\bm{X}$-$\bm{A}$-$\bm{X}'$.
\textbf{b} Blurred RPA results along the same cut, Gaussian broadening chosen to match the DMRG method.
\textbf{c} DMRG results along the same cut, taken from Ref.~\cite{drescher2025}.
\textbf{d-f} Results along the cut $\bm{X}$-$\bm{Y}$-$\bm{K}$.
\textbf{g-i} Results along the cut $\bm{M}$-$\bm{M}'$.
\textbf{j} Blurred RPA results on a $78 \times 6$ cylinder geometry along the $\bm{M}$-$\bm{M}'$ direction.
}
\end{figure}

\clearpage

\begin{figure}[h]
\centering
\includegraphics[scale=0.8]{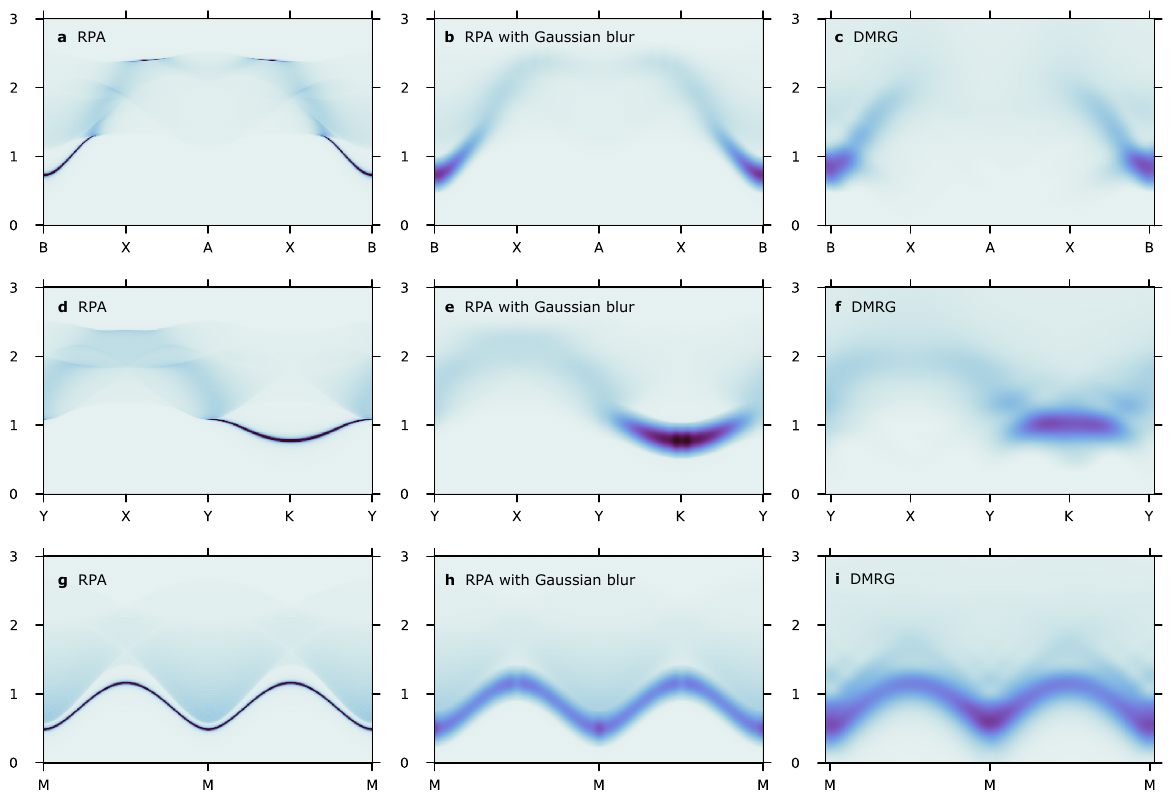}
\caption{\label{fig:dmrg2}
\textbf{Comparison of the RPA method to DMRG results from Ref.~\cite{drescher2025}. Model parameters $J_2/J_1 = 1/8$ and spin-scalar chirality $J_\chi/J_1=0.2$.}
RPA calculations performed on a $L=240$ system. All energies are shown in units of $J_1$.
\textbf{a} RPA results along the direction $\bm{X}$-$\bm{A}$-$\bm{X}'$.
\textbf{b} Blurred RPA results along the same cut, Gaussian broadening chosen to match the DMRG method.
\textbf{c} DMRG results along the same cut, taken from Ref.~\cite{drescher2025}.
\textbf{d-f} Results along the cut $\bm{X}$-$\bm{Y}$-$\bm{K}$.
\textbf{g-i} Results along the cut $\bm{M}$-$\bm{M}'$.
}
\end{figure}

\clearpage

\newpage
\section*{Supplementary Note 3: Gauge fluctuations}
\subsection*{QED$_3$ continuum theory \& CFT}
Coarse graining the Lattice fermions around the Dirac cones, the low energy Lagrangian of the U(1) DSL reads
\begin{equation}
   \mathcal{L}_{\mathrm{QED}_3}=  \sum_i \overline{\psi}_i \left(\slashed{\partial} - \iu \slashed{a} \right) \psi_i + f^{\mu\nu}f_{\mu\nu}
\end{equation}
where $i=1,\dots,N_f$ with $N_f = 4$ fermion flavors.
One can define a topological charge $j_{\mathrm{top}}^\mu = (1/2\pi)\epsilon^{\mu\nu\rho}\partial_\nu a_\rho$ in terms of the emergent gauge field.
Monopoles are local operators which do not conserve this charge, and tunnel $2\pi q$ flux.

The symmetry group of the emergent theory is
\begin{equation}
    \mathrm{SU}(4) \times \mathrm{U}(1)_\mathrm{topo}\simeq \mathrm{SO}(6) \times \mathrm{U}(1)_\mathrm{topo} / \mathbb{Z}_2
\end{equation}
plus discrete symmetries.
The operators in the spectrum can be classified by their transformation under $\mathrm{SO}(6)$ and their topological charge.
Monopoles $\Phi_a$ with $a=1,\dots,6$ transform as the single-index vector representation $\bm{6}$ of $\mathrm{SO}(6)$ and have charge $q=1$.
Their scaling dimension is given by $\Delta_\Phi=1.02$ in the large-$N_f$ expansion \cite{chester2016}, and lies between 1.18--1.34 according to numerical Monte Carlo calculations \cite{karthik2019}.
These six monopoles split up into the three VBS monopoles $\Phi_{a=1,2,3}$ and the three AFM monopoles $\Phi_{a=4,5,6}$, which transform non-trivially under the SO(3)$_{\mathrm{valley}}$ and SO(3)$_{\mathrm{spin}}$ subgroups of emergent SO(6), respectively.
The way that these monopoles transform under the lattice (UV) symmetry group was evaluated in \cite{song2019,song2020}; most importantly, the VBS monopoles transform with momenta $\bm{X}_a=\bm{K}_a/2$, and the AFM monopoles $\bm{K}_{a-3}$.
On the triangular lattice, the first monopole term allowed by lattice symmetries is the triple $q=3$ monopole $\operatorname{Re}\Phi_1\Phi_2\Phi_3$, which is strongly believed to be irrelevant \cite{chester2016}.

\subsection*{Monopole correlation function}

The spin structure factor can be manipulated into the spectral representation by Fourier transforming and then inserting a complete set of states,
\begin{equation}
    S(\bm{q},\omega) = 
    \int \dd{t} \, \eu^{\iu\omega t} \langle {S}_{\bm{q}}^z(t) {S}_{-\bm{q}}^z(0)\rangle
    = 2\pi \sum_n |\langle \Phi_n(\bm{q})| S^z_{\bq}|\Phi_0\rangle|^2 \, \delta(\omega-E_n(\bq)-E_0) \, .
\end{equation}
This correlation function will get contributions from all excited states $\ket{\Psi_n(\bm{q})}$ with appropriate quantum numbers --- namely a time-reversal odd boson with momentum $\bm{q}$.
Up to this point, we have considered only the contribution of fermion degrees of freedom. Next, we must consider the effect of gauge fluctuations, and combine the fermion+gauge contributions consistently.

\subsubsection*{Static correlations}
The relevant gauge excitations for this purpose are the spin-triplet monopoles \cite{song2019,song2020}, which transform as spin operators 
\begin{equation}\label{eq:spin_monopole_generalmap}
    S^\alpha_{\bi} \simeq \alpha_\Phi [\iu  \eu^{\iu \bm{K} \cdot \bm{r}_{\bi}}\,\Phi^\dagger_{\alpha+3}(\bm{r}_{\bi})  + \hc]
\end{equation}
with momentum $\bm{K}$ at the corner of the Brillouin zone. This identification holds up to a magnitude and phase, which we choose to call $\alpha_\Phi$.
If the gauge theory and the spin-triplet monopoles condense $\langle \Phi_{4,5,6}\rangle \neq 0$ (spatially constant) , then this may be identified as an order parameter
\begin{equation}\label{eq:spinmono_orderparam}
    m_{\bi}^\alpha = \langle S^{\alpha}_{\bi} \rangle \simeq \alpha_\Phi \operatorname{Re}[\iu \eu^{\iu \bm{K} \cdot \bm{r}_{\bi}}\,\langle\Phi^\dagger_{\alpha+3}\rangle]
\end{equation}
which oscillates between the three sublattices of the 120-degree ordered state.
In our case, we will not condense this order parameter but use \eqref{eq:spin_monopole_generalmap} to identify the quasi-long-range correlations via
\begin{equation}\label{eq:spincorr_mapping}
    \langle S_{\bi}^\alpha S_{\bj}^\alpha \rangle \simeq\alpha_\Phi ^2
    \operatorname{Re}[ \eu^{\iu \bm{K} \cdot (\bm{r}_{\bi}-\bm{r}_{\bj})}
    \langle \Phi^\dagger(\bm{r}_{\bi})\Phi(\bm{r}_{\bj})\rangle]
\end{equation}
where we take $\Phi = \Phi_6$ to be one of the SO(3)-spin monopoles.
Importantly, this correlation function is taken at equal times.
We know from the CFT results that this correlation function is completely fixed as a power law; when the continuum theory holds
\begin{equation}\label{eq:monopole_corrfn_cft_app}
    \langle \Phi^\dagger(\bm{r}_{\bi})\Phi(\bm{r}_{\bj})\rangle \sim |\bm{r}_{\bi} - \bm{r}_{\bj}|^{-2\Delta_\Phi} \, ,
\end{equation}
where $\Delta_\Phi$ is the monopole scaling dimension.
We bring together Eqns.~\eqref{eq:spincorr_mapping} and \eqref{eq:monopole_corrfn_cft_app} to get
%
%
\begin{equation}
    \langle S^z_{\bm{i}} S^z_{\bj}\rangle
    \sim 
     \real{\left[ \eu^{\iu \bm{K} \cdot (\bm{r}_{\bi}-\bm{r}_{\bj})}
    \frac{\alpha_\Phi^2}{|\bm{r}_{\bi} - \bm{r}_{\bj}|^{2\Delta_\Phi}}\right]}
    \,\,
\end{equation}
The correlation function $\langle S^z_{\bm{i}} S^z_{\bj}\rangle$ is algebraic with an oscillating phase $\cos(\bm{K}\cdot (\bm{r}_{\bi}-\bm{r}_{\bj}))$. 
In this work, we take the numerical value $\Delta_\Phi=1.25$ when needed, compatible with numerical estimates which put it in the range $1.18$--$1.34$ \cite{karthik2019}.

\subsubsection*{Dynamic correlations}

We can also calculate the contribution of monopoles to the dynamical susceptibility. We substitute Eq.~\eqref{eq:spin_monopole_generalmap} into the time-ordered correlation function
\begin{align}
    \chi_\Phi(\bm{q},\iu\omega_n) & =
    \frac{1}{N}\sum_{\bi\bj} \eu^{-\iu \bm{q}\cdot (\bm{r}_{\bi} - \bm{r}_{\bj})}  \int \dd{\tau} \, \eu^{-\iu\omega_n \tau} \langle T_\tau \, {S}_{\bi}^z(\tau) {S}_{\bj}^z(0)\rangle\,, \\
    &=  \frac{\alpha_\Phi ^2}{N}\sum_{\bi\bj} \eu^{-\iu \bm{q}\cdot (\bm{r}_{\bi} - \bm{r}_{\bj})} \int \dd{\tau} \, \eu^{\iu\omega \tau} \eu^{\iu \bm{K} \cdot (\bm{r}_{\bi}-\bm{r}_{\bj})}
    \langle \Phi^\dagger(\bm{r}_{\bi},\tau)\Phi(\bm{r}_{\bj},0)\rangle\,.
\end{align}
Using translational invariance, we perform the sum over $\bj$ to yield
\begin{align}
    \chi_\Phi(\bm{q},\iu\omega_n) & = \alpha_\Phi ^2 
    \sum_{\bi} \int \dd{\tau} \,  \eu^{-\iu (\bm{q}-\bm{K})\cdot \bm{r}_{\bi}-\iu \omega_n \tau} \langle \Phi^\dagger(\bm{r}_{\bi},\tau)\Phi(\bm{0},0)\rangle\,.
\end{align}
We can change variables to $x=(\tau, \bm{r}_{\bi} - \bm{r}_{\bj})$ and $q=(\omega_n,\bm{K}-\bm{q})$ to make the resulting problem explicit; the answer is the Fourier transform of a power law
\begin{align}
       \chi_\Phi(\bm{q},\iu\omega_n) &= \alpha_\Phi^2\int \dd[3]{x} \, \eu^{-\iu q\cdot x} 
    \langle \Phi^\dagger(x)\Phi(0)\rangle \\
    &= \alpha_\Phi^2\int\dd[3]{x} \eu^{-\iu q\cdot x} |x|^{-2\Delta_\Phi}
    = c \alpha_\Phi^2|q|^{2\Delta_\Phi-3} \,,
\end{align}
where $c = \pi^{2\Delta_\Phi-3/2}\Gamma(3/2-\Delta_\Phi)/\Gamma(\Delta_\Phi)$ when $0<\Delta_\Phi < 3/2$.
The result is hence another power law; in terms of the components it tells us
\begin{equation}
     \chi_\Phi(\bm{q},\iu\omega_n) = \frac{c \alpha_\Phi^2}{(|\bm{K}-\bq| + \omega_n^{\, 2})^{3/2-\Delta_\Phi}} \, 
\end{equation}
and finally transforming back into real-time (when $\omega > 0$) we recover the spin susceptibility due to the monopoles
\begin{fmffile}{monopole-1}
\begin{equation}\label{eq:monosusc_th}
    \chi_\Phi(\bm{q}, \omega+\iu \eta) = {\frac{c\alpha_\Phi^2}{(|\bm{K}-\bm{q}|^2-\omega^2 -\iu \eta)^{3/2-\Delta_\Phi}}}
    \,\,
=
\,\,
S^z\,
\begin{gathered}
\begin{fmfgraph*}(50,10)
    \fmfleft{i1}
    \fmfright{o1}
    \fmf{boson}{i1,o1}
\end{fmfgraph*}
\end{gathered}
\, S^z
\end{equation}
\end{fmffile}
In this Feynman diagram, we represent the real-space monopole propagator as a wiggly line.
By taking the imaginary part, this produces the contribution of monopoles to the dynamical structure factor \cite{hermele2008,seifert2024}.
\begin{align}
    S(\bm{q}, \omega) &= \imaginary{\chi_\Phi(\bm{q}, \omega+\iu \eta)} 
    = -\imaginary{\chi_\Phi(\bm{q}, \omega-\iu \eta)} 
    \\
    &= \imaginary{\left[{\frac{c\alpha_\Phi^2}{(|\bm{K}-\bm{q}|^2-\omega^2 +\iu \eta)^{3/2-\Delta_\Phi}}}\right]}\, .
    \label{eq:monopolesusc_app}
\end{align}
This result resembles the spectral function across quantum critical transitions in one dimension where the quasiparticle picture breaks down. It is also equivalent to the result for the Luttinger liquid when one replaces the exponents $3-2\Delta_\Phi \to 1$.

Taking the imaginary part, we find
\begin{equation}
    S(\bq,\omega) 
    = \theta(\omega^2-\delta q^2) \frac{ c \, \sin[\pi(3/2-\Delta_\Phi)]}{|\omega^2-\delta q^2 |^{3/2-\Delta_\Phi}} \, .
\end{equation}
At $\delta q = 0$, this diverges as
\begin{equation}
    S(\bm{K},\omega) \sim \omega^{2\Delta_\Phi-3} \, .
\end{equation}
The lightcone structure is enforced by the step function, and the divergence as $\omega\to|\delta q|$ is $\sim |\omega-\delta q|^{\Delta-3/2}$.

This result is derived using the limit of large distances $r\gg a$, where the continuum theory holds. It consequently describes the limiting behavior as $\omega\to0$.
At lattice scale distances, the monopole (diverging) correlation function must be cut-off by a UV regulator.
In the corresponding regime $\omega\sim 1$, we therefore suppress the correlation function, such that its Fourier transform converges as $r\to 0$. We implement this simply using a Gaussian cutoff in our calculations of the monopole susceptibility
\begin{equation}
        S(\bm{q}, \omega)_{\mathrm{lattice}} = S(\bm{q},\omega) \exp(-\omega^2/\omega_{\mathrm{l}}^2)
\end{equation}
where we take $\omega_{\mathrm{l}} = 0.5 \, J_1$.

\subsection*{Monopole-paramagnon hybridization}

Both monopoles and spinons contribute to the structure factor. But how do these two different excitations interact with each other?
Next, we will put together these parts and derive the hybridization of these  excitations.

The challenge will be to work out the effect of the quantum-critical continuum at the $\bm{K}$ point on the response of the spinon exciton mode.
When adding these contributions, it is worth
noting that although the fermion bilinears and monopoles both transform like spin operators, their normalization is different.
In the case of spinons, the normalization is `fixed' by the Abrikosov decomposition.
The monopoles however are not defined as some local quasiparticle in this basis, and as such we can only write $S^z_{\bi} \simeq \Phi({\bm{r}_{\bi}})$ as in \eqref{eq:spin_monopole_generalmap} up to some arbitrary normalization (and a phase fixed by symmetry). 
We choose the following normalization for the two types of operators
\begin{align}
    S_{\bi}^z &= \frac{1}{2}f^\dagger_{\bi}\sigma^z f_{\bi}^\dagga \\  
    S_{\bi}^z &= \alpha_{\Phi} \left[ \iu \eu^{\iu\bm{K}\cdot \bm{r}_{\bi}}\Phi^\dagger(\bm{r}_{\bi}) + \hc \right]
\end{align}
We initially describe the spin-spin correlation function as a sum of the fermion and gauge contributions, starting with a non-interacting fermion theory.
In diagrammatic language, we write the mean-field susceptibility including gauge fluctuations as
\begin{fmffile}{combsusc_freecontribution}
\begin{equation}\label{eq:app_twocontr_sqw}
\chi(\bm{q},\omega)
=\chi_f^0(\bm{q},\omega)+\alpha_\Phi^2\,\chi_\Phi(\bm{q},\omega)
\,\,
=
\,\,
\begin{gathered}
\begin{fmfgraph*}(33,30)
 \fmfleftn{l}{1}\fmfrightn{r}{1}
 \fmf{fermion,right=.5}{l1,r1}\fmf{fermion,right=.5}{r1,l1}
\end{fmfgraph*}
\end{gathered}
\,\,
+
\,\,
\begin{gathered}
\begin{fmfgraph*}(50,10)
    \fmfleft{i1}
    \fmfright{o1}
    \fmf{boson}{i1,o1}
\end{fmfgraph*}
\end{gathered}
\end{equation}
\end{fmffile}\noindent
%
The monopole susceptibility Eq.~\eqref{eq:monopolesusc_app}, and hence also the relative magnitude of the terms in Eq.~\eqref{eq:app_twocontr_sqw}, is defined up to a constant $\alpha_\Phi^{\, 2}$.


In order to describe the effect of gauge fluctuations, we will now write down a fermion-monopole coupling and treat its effect perturbatively alongside the fermion-fermion interactions. 
In real space, this coupling is simply a coupling between the spin-order parameters of the fermions with wavevector $\bm{K}$, defined as
\begin{equation}
      S^\alpha_{\bi}  = m_{\bi}^\alpha =\eu^{\iu\bm{K}\cdot \bm{r}_{\bi}} n_{\bi}^\alpha + \mathrm{c.c.} \, ,
\end{equation}
and the monopole \eqref{eq:spinmono_orderparam}:
\begin{equation}
    g \alpha_\Phi [\iu \Phi^\dagger(\bm{r}_{\bi}) n^z_{\bi} + \hc ]\, .
\end{equation}
In the fermionic parton representation, this leads to the vertex
\begin{equation}
         = \frac{1}{2} g\alpha_\Phi
         \left( \iu \Phi^\dagger \eu^{\iu\bm{K}\cdot \bm{r}_{X}}\sum_{\bk}  f^\dagger_{\bm{k},X}\sigma^z f^\dagga_{X,\bm{k}+\bm{K}_\alpha} + \hc\right)
     =
 \begin{fmffile}{monopolefermion-vertex}
\begin{gathered}
    \begin{fmfgraph*}(30,30)
    \fmfleft{i1}\fmfrightn{o}{2}
    \fmf{boson}{i1,w1}
    \fmf{fermion,tension=.5}{w1,o1}\fmf{fermion,tension=.5}{o2,w1}
\end{fmfgraph*}
\end{gathered}
\end{fmffile}
\end{equation}
This result is written in the sublattice basis, where the phase $R_X = \eu^{\iu\bm{K}\cdot \bm{r}_{X}}$ comes from the relative position of the site $X$ in the unit cell.

In diagrams, we replace the free fermion bubble $\chi_f^0(\bq,\omega)$ in Eq.~\eqref{eq:app_twocontr_sqw} by the interacting fermion susceptibility $\chi_f(\bq,\omega)$
\begin{fmffile}{combsusc_inter}
\begin{equation}
\chi(\bm{q},\omega)
=\chi_f(\bm{q},\omega)+ \chi_\Phi(\bm{q},\omega)
\,\,
=
\,\,
\begin{gathered}
\begin{fmfgraph*}(60,30)
 \fmfleftn{l}{1}\fmfrightn{r}{1}
\fmfrpolyn{shaded}{G}{4}
 \fmf{fermion,right=0.25,tension=2}{l1,G1}\fmf{fermion,right=0.25,tension=2}{G2,l1}
 \fmf{fermion,right=0.25,tension=2}{r1,G3}\fmf{fermion,right=0.25,tension=2}{G4,r1}
\end{fmfgraph*}
\end{gathered}
\,\,
+
\,\,
\begin{gathered}
\begin{fmfgraph*}(50,10)
    \fmfleft{i1}
    \fmfright{o1}
    \fmf{boson}{i1,o1}
\end{fmfgraph*}
\end{gathered}
\end{equation}
\end{fmffile}
The diagrams which correct the bare susceptibility to leading order are as follows
\begin{fmffile}{susceptibilityseries_monopole}
\begin{multline}
\begin{gathered}
\begin{fmfgraph*}(60,30)
 \fmfleftn{l}{1}\fmfrightn{r}{1}
\fmfrpolyn{shaded}{G}{4}
 \fmf{fermion,right=0.25,tension=2}{l1,G1}\fmf{fermion,right=0.25,tension=2}{G2,l1}
 \fmf{fermion,right=0.25,tension=2}{r1,G3}\fmf{fermion,right=0.25,tension=2}{G4,r1}
\end{fmfgraph*}
\end{gathered}
\,\,
=
\,\,
\begin{gathered}
\begin{fmfgraph*}(33,30)
 \fmfleftn{l}{1}\fmfrightn{r}{1}
 \fmf{fermion,right=.5}{l1,r1}\fmf{fermion,right=.5}{r1,l1}
\end{fmfgraph*}
\end{gathered}
\,\,
+
\,\,
\begin{gathered}
\begin{fmfgraph*}(80,30)
 \fmfleftn{l}{1}\fmfrightn{r}{1}
  \fmfdot{K1}\fmfdot{K2}
 \fmf{fermion,fermion,right=.5}{l1,K1}\fmf{fermion,fermion,right=.5}{K1,l1}
 \fmf{dashes,tension=4}{K1,K2}
 \fmf{fermion,fermion,right=.5}{r1,K2}\fmf{fermion,fermion,right=.5}{K2,r1}
\end{fmfgraph*}
\end{gathered}
\,\,
+
\,\,
\begin{gathered}
\begin{fmfgraph*}(80,30)
 \fmfleftn{l}{1}\fmfrightn{r}{1}
 \fmf{fermion,fermion,right=.5}{l1,K1}\fmf{fermion,fermion,right=.5}{K1,l1}
 \fmf{boson,tension=4}{K1,K2}
 \fmf{fermion,fermion,right=.5}{r1,K2}\fmf{fermion,fermion,right=.5}{K2,r1}
\end{fmfgraph*}
\end{gathered}
\\
\,\,+\,\,
\begin{gathered}
\begin{fmfgraph*}(125,30)
 \fmfleftn{l}{1}\fmfrightn{r}{1}
  \fmfdot{K1}\fmfdot{K2}
 \fmf{fermion,fermion,right=.5}{l1,K1}\fmf{fermion,fermion,right=.5}{K1,l1}
 \fmf{dashes,tension=4}{K1,K2}
   \fmfdot{K3}\fmfdot{K4}
 \fmf{fermion,fermion,right=.5}{K2,K3}\fmf{fermion,fermion,right=.5}{K3,K2}
 \fmf{dashes,tension=4}{K3,K4}
 \fmf{fermion,fermion,right=.5}{r1,K4}\fmf{fermion,fermion,right=.5}{K4,r1}
\end{fmfgraph*}
\end{gathered}
\,\,+\,\,
\begin{gathered}
\begin{fmfgraph*}(125,30)
 \fmfleftn{l}{1}\fmfrightn{r}{1}
 \fmf{fermion,fermion,right=.5}{l1,K1}\fmf{fermion,fermion,right=.5}{K1,l1}
 \fmf{boson,tension=4}{K1,K2}
 \fmf{fermion,fermion,right=.5}{K2,K3}\fmf{fermion,fermion,right=.5}{K3,K2}
 \fmf{boson,tension=4}{K3,K4}
 \fmf{fermion,fermion,right=.5}{r1,K4}\fmf{fermion,fermion,right=.5}{K4,r1}
\end{fmfgraph*}
\end{gathered}
\\
\,\,+\,\,
\begin{gathered}
\begin{fmfgraph*}(125,30)
 \fmfleftn{l}{1}\fmfrightn{r}{1}
  \fmfdot{K1}\fmfdot{K2}
 \fmf{fermion,fermion,right=.5}{l1,K1}\fmf{fermion,fermion,right=.5}{K1,l1}
 \fmf{dashes,tension=4}{K1,K2}
 \fmf{fermion,fermion,right=.5}{K2,K3}\fmf{fermion,fermion,right=.5}{K3,K2}
 \fmf{boson,tension=4}{K3,K4}
 \fmf{fermion,fermion,right=.5}{r1,K4}\fmf{fermion,fermion,right=.5}{K4,r1}
\end{fmfgraph*}
\end{gathered}
\,\,+\,\,
\begin{gathered}
\begin{fmfgraph*}(125,30)
 \fmfleftn{l}{1}\fmfrightn{r}{1}
 \fmf{fermion,fermion,right=.5}{l1,K1}\fmf{fermion,fermion,right=.5}{K1,l1}
 \fmf{boson,tension=4}{K1,K2}
   \fmfdot{K3}\fmfdot{K4}
 \fmf{fermion,fermion,right=.5}{K2,K3}\fmf{fermion,fermion,right=.5}{K3,K2}
 \fmf{dashes,tension=4}{K3,K4}
 \fmf{fermion,fermion,right=.5}{r1,K4}\fmf{fermion,fermion,right=.5}{K4,r1}
\end{fmfgraph*}
\end{gathered}
\,\,+\,\,\cdots
\end{multline}
\end{fmffile}
Beyond leading order, the series of diagrams will contain all combinations of Heisenberg interactions and monopole propagators.
The Dyson equation is modified to include both intermediate states:
\begin{fmffile}{susceptibilityresum_mono}
\begin{equation}
%
\begin{gathered}
\begin{fmfgraph*}(60,30)
 \fmfleftn{l}{1}\fmfrightn{r}{1}
\fmfrpolyn{shaded}{G}{4}
 \fmf{fermion,right=0.25,tension=2}{l1,G1}\fmf{fermion,right=0.25,tension=2}{G2,l1}
 \fmf{fermion,right=0.25,tension=2}{r1,G3}\fmf{fermion,right=0.25,tension=2}{G4,r1}
\end{fmfgraph*}
\end{gathered}
\,\,
=
\,\,
\begin{gathered}
\begin{fmfgraph*}(33,30)
 \fmfleftn{l}{1}\fmfrightn{r}{1}
 \fmf{fermion,right=.5}{l1,r1}\fmf{fermion,right=.5}{r1,l1}
\end{fmfgraph*}
\end{gathered}
\,\,
+
\,\,
\begin{gathered}
\begin{fmfgraph*}(110,30)
 \fmfleftn{l}{1}\fmfrightn{r}{1}
  \fmfdot{K1}\fmfdot{K2}
 \fmf{fermion,fermion,right=.5,tension=1.5}{r1,K1}\fmf{fermion,fermion,right=.5.,tension=1.5}{K1,r1}
 \fmf{dashes,tension=6}{K1,K2}
 \fmfrpolyn{shaded}{G}{4}
 \fmf{fermion,right=0.25,tension=2}{K2,G1}\fmf{fermion,right=0.25,tension=2}{G2,K2}
 \fmf{fermion,right=0.25,tension=2}{l1,G3}\fmf{fermion,right=0.25,tension=2}{G4,l1}
\end{fmfgraph*}
\end{gathered}
\,\,
+
\,\,
\begin{gathered}
\begin{fmfgraph*}(110,30)
 \fmfleftn{l}{1}\fmfrightn{r}{1}
 %
 \fmf{fermion,fermion,right=.5,tension=1.5}{r1,K1}\fmf{fermion,fermion,right=.5.,tension=1.5}{K1,r1}
 \fmf{boson,tension=6}{K1,K2}
 \fmfrpolyn{shaded}{G}{4}
 \fmf{fermion,right=0.25,tension=2}{K2,G1}\fmf{fermion,right=0.25,tension=2}{G2,K2}
 \fmf{fermion,right=0.25,tension=2}{l1,G3}\fmf{fermion,right=0.25,tension=2}{G4,l1}
\end{fmfgraph*}
\end{gathered}
\end{equation}
\end{fmffile}
This can be formulated once again as a matrix equation \eqref{eq:dyson_rpa}, where the interaction now includes two terms: the Heisenberg four-fermion vertex, and an intermediate-monopole propagator
\begin{equation}
    \hat{U}(\bq,\omega)^{\mu\nu} = \delta^{ij}[-2J_{\bm{q}}^{XY} + (g^2/4) (RR^\dagger)^{XY} \chi_{\Phi}(\bq,\omega)] \, .
\end{equation}
The relative phases are $(RR^\dagger)^{XY} = \eu^{\iu\bm{K}\cdot (\bm{r}_{X}-\bm{r}_{Y})}$.
The Dyson equation for the fermion susceptibility in this language is exactly as before
\begin{align}\label{eq:dyson_rpa2}
    [\chi_f(\bq,\omega)]^{\mu\nu} &= [\chi_f^0(\bq,\omega)]^{\mu\nu} + [\chi_f(\bq,\omega)]^{\mu\rho}[U(\bq,\omega)]^{\rho\sigma}[\chi_f^0(\bq,\omega)]^{\sigma\nu} \, ;
\end{align}
resumming this series using the RPA produces the susceptibility in the presence of fermion-fermion and fermion-gauge interactions.
This is the method used to calculate the interacting susceptibility in the main text.
In that case, we take the numerical values $\alpha_\Phi=0$ and $g = 0.05$.


\subsection*{Continuum theory of transitions}\label{sec:cont_transitions_theory}


\subsubsection*{Stripe transition}

In this special case, the 2-site unit cell of stripe order is commensurate with the unit cell of the QSL Ansatz.
The Hubbard--Stratonovich field $\phi^i_{\bm{M},X}$ therefore transforms like a fermion mass in the low-energy description. This is compatible with symmetry-based field-theory works \cite{song2019,song2020}. 
%
%
In the continuum theory, the condensing boson couples to the SU(4) fermions via a mixed chiral Gross--Neveu coupling
\begin{equation}\label{eq:qedgn_fermions}
   \mathcal{L}_{\mathrm{cGN}} =  -h \sum_j \left[(\overline{\psi} \vec{\sigma} \tau^j \psi) \cdot \vec{\varphi}^j
   +  m^2 \,|\vec{\varphi}^j|^2 \right].
\end{equation}
The boson mass $m^2$ is defined within the Landau--Ginzburg theory in the Methods of the main text.
The overall action is given by $\mathcal{L}_{\mathrm{QED}_3} + \mathcal{L}_{\mathrm{cGN}}$.
The boson represents the fluctuating magnetic order at the $\bm{M}$ point, and naturally wants to precipitate stripe order when it condenses at $m^2=0$.
When the spinon exciton condenses, it will also precipitate a mass for the fermions in some channel $M_{ij}$.
This confines the gauge field by the Polyakov mechanism, which will in turn proliferate 120-degree or VBS order in the IR limit.
The exact nature of the competing order depends on which of the chiral masses $M_{ij}$ the interaction \eqref{eq:qedgn_fermions} selects.
%

We believe this picture will be modified when looking at the energetics of the competing magnetic orders around this transition point.
Our calculations suggest this continuous transition would first occur when $r_c^{\mathrm{stripe}} = 0.18$. 
We know that stripe order is energetically preferred to 120-degree order for all $r>0.125$, even at the classical level.
We suggest instead that this continuous transition at $r=0.18$ should be preempted by a \emph{first-order} transition to stripe magnetic order. 
This perspective is compatible with the numerical works on this model which consistently see a first-order transition at lower $J_2/J_1$ \cite{kaneko2014,hu2015,zhu2015,iqbal2016,hu2019,ferrari2019,drescher2023}.
Our method is not able to compare energies between liquid and ordered states, but we can provide an upper bound to the regime of stability of a Dirac spin liquid in a lattice model.
It is an open question how to understand the position of this first-order transition analytically. 

\subsubsection*{120-degree transition}

Next, we describe the effect of gauge fluctuations on the transition at low $J_2/J_1$.
In contrast to the stripe transition, we believe this will realise a continuous transition in the presence of gauge fluctuations. We will show this is in a QED$_3$ O(3)-Gross--Neveu universality class by studying the low-energy continuum theory.

This transition occurs as the mass of the spinon-exciton mode goes to zero at the $\bm{K}$ point.
Here, the relevant symmetry-compatible operators are not fermion bilinears as in the typical Gross--Neveu--Yukawa models \cite{janssen2017,song2019,dupuis2019,dupuis2022}, but rather the relevant spin-triplet monopoles.
The effective continuum theory reads
\begin{equation}\label{eq:qedgn_monos}
   \mathcal{L}_{\Phi\mathrm{GN}} =  g \sum_\alpha \left(\iu{\Phi}_{\alpha+3}^\dagger \phi_\alpha + \hc\right)+ m^2|\vec{\phi}|^2 \, ,
\end{equation}
where $\vec{\phi} = \vec{\phi}_{\bm{K}}$ here, and $m^2=m^2_\phi(\bm{K})\sim (r-r_c^{120})$ from the Landau--Ginzburg theory. 
We call this a monopole-Gross--Neveu ($\Phi$GN) coupling.
The total theory is given by $\mathcal{L}_{\mathrm{QED}_3}+ \mathcal{L}_{\Phi\mathrm{GN}}$.
Integrating out the fluctuating bosons produces an interaction term
\begin{equation}\label{eq:gnphi123}
       \mathcal{L}_{\mathrm{QED}_3} + (g/m)^2 \left( 
       \Phi_4^\dagger\Phi^\dagga_4 + \Phi_5^\dagger\Phi^\dagga_5 + \Phi_6^\dagger\Phi^\dagga_6\right) \,.
\end{equation}
What is the nature of the resulting critical point driven by $m^2\to 0$?
As we introduced above, the OPE of two monopoles $\Phi^\dagger\Phi$ produces charge-zero operators.
The symmetric combination in \eqref{eq:gnphi123} specifically leads to only four-fermion interactions appearing.
The most dominant OPE channel from the above effective monopole interaction \eqref{eq:gnphi123} is the four-fermion interaction in the irrep $\bm{20}'$ of SU(4)
\begin{equation}\label{eq:chgnop}
   \mathcal{O}_{\Phi\mathrm{GN}} \sim (\overline{\psi}_i\vec{\sigma}\psi_i)^2 
   \, .
\end{equation}
This interaction is believed to be irrelevant, but will cause a phase transition as the spinon exciton condenses at $m^2=0$.
The interaction \eqref{eq:chgnop} is a chiral Heisenberg Gross--Neveu interaction.
We predict it will lead to the same fixed-point as the chiral Heisenberg Gross--Neveu--Yukawa model described by Ref.~\cite{dupuis2019} on the kagome lattice.
We therefore suggest that the spin liquid--120-degree order transition at $J_{2,c}/J_1 \approx 0.07$ on the triangular lattice is in this same universality class.

\subsection*{Critical behavior}

\subsubsection*{Critical point}

{The chiral Heisenberg Gross--Neveu-QED$_3$ critical point describes the end point of a stable QED$_3$ CFT, driven through a phase transition by large fermion-fermion interactions \cite{dupuis2019}. 
It has a reduced $\mathrm{SO}(3)\times \mathrm{SO}(3)$ symmetry group compared to the $\mathrm{SO}(6)$ of the original model, since these interactions only couple the spin components of the fermion masses.
The degeneracy of the spin-triplet and spin-singlet monopoles is broken by the interactions, which tend to condense the spin-triplet order.
In our model, the same is achieved via monopole-monopole terms~\eqref{eq:gnphi123} which only act in the spin-triplet subgroup.
This results in a hierarchy of different monopole scaling dimensions;
the most relevant are the spin-triplet monopoles, with scaling dimensions
\begin{equation}
  \Delta^{\mathrm{GN}}_{S=1} = 0.195 \, N_f + \mathcal{O}(N_f^0)
\end{equation}
in a large-$N_f$ expansion \cite{dupuis2019}. 
The spin-singlet monopoles have the dimension
\begin{equation}
  \Delta^{\mathrm{GN}}_{S=0} = \Delta^{\mathrm{GN}}_{S=1} + 0.546 + \mathcal{O}(N_f^{-1}) \, .
\end{equation}
In our $N_f=4$ theory of the U(1) Dirac QSL, we find $\Delta^{\mathrm{GN}}_{S=1} \approx 0.78$ and $\Delta^{\mathrm{GN}}_{S=0} = 1.33$ \cite{dupuis2019}. Contrast this to the (degenerate) value $\Delta_\Phi = 1.02$ in standard QED$_3$ \cite{pufu2014}.

Let us turn now to the physical consequences of these exponents for our model.
The spin structure factor at the new critical point will take the same form as Eq.~\eqref{eq:monosusc_th}, but with a modified exponent
\begin{equation}
      \chi_{\mathcal{S}}(\bm{q}, \omega+\iu \eta)
    = \frac{c}{(|\bm{K}-\bm{q}|^2-\omega^2 -\iu \eta)^{3/2-{\Delta}^{\mathrm{GN}}_{S=1}}}
    \label{eq:monopolesusc_qedgn}
\end{equation}
due to the shifted spin-triplet monopole scaling dimension.
The lower value of $\Delta_\Phi$ represents a stronger divergence and hence a higher spectral weight in $S(\bq,\omega)$ at low energies. This is consistent with the spinon exciton coming down and transferring spectral weight towards zero. 

Finally, we know the transition is driven by tuning $m^2=r_c^{120}-r$. 
This coupling is the bosonic mass.
Its scaling dimension was also calculated in the large-$N_f,N_b$ limit of the (noncompact) QED$_3$-cHGN model in $d=4-\epsilon$ dimensions \cite{dupuis2019}, giving
$\Delta_{\varphi^2} = 1+ 4.577\epsilon$ when $N_f=4$.
The Lagrangian for this coupling is $m^2\varphi^2$ and as such the coupling has scaling dimension $3-\Delta_{\varphi^2}$. This means that in the critical regime, the correlation length will obey
\begin{equation}
  \xi^{3-\Delta_{\varphi^2}} \sim |r_c^{120}-r|^{-1} \, .
\end{equation}
Hence, we can define the correlation length exponent
\begin{equation}\label{eq:chi_qed3gn}
  \xi\sim |r_c^{120}-r|^{-\nu}
\end{equation}
with $\nu^{-1}=3-\Delta_{\varphi^2}$.
Setting $\epsilon=1$, we finally find that $\nu = 1.644$ in our model.\footnote{The value of $\nu$ depends upon how the $4-\epsilon$ expansion is taken \cite{dupuis2019}.}
Equivalent to Eq.~\eqref{eq:chi_qed3gn}, there is a mass that scales $m\sim\xi^{-1} \sim |r_c^{120}-r|^{\nu}$. In the Gaussian theory this was $\nu=1/2$.
}

\subsubsection*{Scaling collapse}

{When the  the gap to the spinon exciton is small, the perturbative RPA approach will break down as the corrections diverge.
Returning to Fig.~4\textbf{b} of the main text, we now turn to the behavior of the spectral function when the mode approaches the critical point.
When $J_2=J_{2,c}$ we have a gapless boson $m^2=0$ and we sit at the new QED$_3$-cHGN fixed point. Here, we apply the Ansatz~\eqref{eq:monopolesusc_qedgn} to show $S(\bm{K},\omega) \sim \omega^{2\Delta^{\mathrm{GN}}_{S=1} - 3}$. This shifted exponent is plotted as the critical line in Fig.~4\textbf{b}.

If, instead, we sit just away from the critical point, then we know that the gap scales as $m\sim (r-r_c^{120})^{1.6}$ in the large-$N_f$ and $4-\epsilon$ approximations.
Instead, here we can write a scaling Ansatz for the critical regime.
To do this, first use that when $\omega \ll m$, the effective continuum theory is just QED$_3$. The exponents in this regime should be dictated by $\Delta_\Phi=1.25$.
Alternatively, when $\omega \gg m$, the spectral function should scale with QED$_3$-GN exponents, since at these energy scales it looks like the critical point $m=0$.
In the intermediate regime we can write down a scaling Ansatz
\begin{equation}\label{eq:crossover_ansatz}
  S(\bm{K},\omega) = \omega^{2\Delta_\Phi-3} \mathcal{G}(\omega / m) \, ,
\end{equation}
where $\mathcal{G}(x)$ is constrained by the two limiting cases discussed above.
A basic example is plotted as the crossover line in Fig.~4\textbf{b}, where we take $\mathcal{G}(x) = (c+x^s)^{(\Delta^{\mathrm{GN}}_{S=1} - \Delta_\Phi)/s}$, where we take $s=2.5$ and $c$ is a complex constant.
It recovers both limits as discussed, and has a crossover between power laws around $m=0.05$.
Imagining a numerical simulation or experiment where the interaction strength $r$ could be tuned, this would allow a scaling collapse to be obtained and extract the relevant critical exponents.
Combined with the scaling properties of $m$, we would expect a scaling collapse by plotting $\omega^{3-2\Delta_\Phi} S(\bm{K},\omega)$ against $|r-r_c^{120}|^{\nu_{\mathrm{GN}}} \omega$.

We may also evaluate the real-space spin-spin correlations $g_{\bij}$ within these three regimes, given by Eq.~\eqref{eq:gbij}.
Away from the critical point, the long-distance behavior is dominated by a power-law tail, governed by the universal QED$_3$ monopole exponents $g(r)\sim r^{-2\Delta_\Phi}$. 
The finite-energy spinon exciton causes a non-universal correction at short distances, with a characteristic energy scale that goes as $\xi \sim (r-r_c^{120})^{-1/2}$.
Numerical simulations suggest the monopole contribution is subleading to the fermionic one (cf. Supplementary Fig.~\ref{fig:realspace_rpa}), but it must still dominate for large separations.
Approaching the critical point, this correlation length will diverge, governed by new cHGN exponents $\xi\sim (r-r_c^{120})^{-\nu_{\mathrm{GN}}}$.
For lengths $r\ll\xi$, the correlations will be controlled by cHGN exponents $g(r)\sim r^{-2\widetilde{\Delta}_\Phi}$ such that at the critical point $\xi\to\infty$ we have a picture of conformal correlations.
Away from the transition, the correlations for lengths $r\gg\xi$ will be given by QED$_3$ exponents.
We equally expect a scaling collapse by plotting $r^{2\Delta_\Phi} g(r)$ against $|r-r_c^{120}|^{-\nu_{\mathrm{GN}}} \omega$.
}

\newpage
\section*{Supplementary Note 4: Chiral Spin Liquid}

\begin{figure}
\centering
\includegraphics[width=0.6\textwidth]{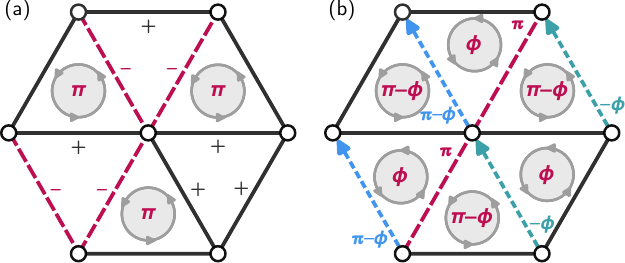}
\caption{{Dirac and chiral spin liquid Ans\"atze.} (a) The $[\pi,0]$ Ansatz of the U(1) Dirac QSL. (b) The general $[\pi-\phi,\phi]$ Ansatz. In this work, it is generated by a self-consistent mean-field treatment of a spin-scalar chirality interaction which breaks time reversal symmetry.\label{fig:chiralansatz}}
\end{figure}

\subsection*{Chiral QSL on the triangular lattice}
In the main text, we introduced the spin-scalar chirality (SSC) interaction~\cite{hu2016variational}
\begin{equation}\label{eq:h12chi}
    \mathcal{H} = \mathcal{H}_{J_1J_2} + J_\chi\! \sum_{\bi\bj\bk \in \triangle,\bigtriangledown} \!\!\! \vec{S}_{\bi}\cdot (\vec{S}_{\bj}\cross \vec{S}_{\bk})\, ,
\end{equation}
where $\bi\bj\bk$ is an ordered set of the three spins on each triangular plaquette.
The Hamiltonian Eq.~\eqref{eq:h12chi} takes the following form \cite{banerjee2023,maity2025} 
\begin{multline}\label{eq:H0chiraldecomp}
  \mathcal{H}_0 = -\frac{1}{2} \sum_{\bi\bj}J_{\bij} [(\chi_{\bij} f^\dagger_{\bj\alpha} f^\dagga_{\bi\alpha} + \hc) - |\chi_{\bij}|^2] 
  + \frac{3\iu }{4} \sum_{\bi\bj\bk} J_\chi (
  \chi_{\bi\bk}\chi_{\bk\bj} f^\dagger_{\bj\alpha} f^\dagga_{\bi\alpha} + 
  \chi_{\bi\bk}\chi_{\bj\bi} f^\dagger_{\bk\alpha} f^\dagga_{\bj\alpha}
  \\
  +\chi_{\bk\bj}\chi_{\bj\bi} f^\dagger_{\bi\alpha} f^\dagga_{\bk\alpha}
  - \chi_{\bi\bk}\chi_{\bk\bj}\chi_{\bj\bi}
   - \hc) \, .
\end{multline}
Assuming that $J_\chi/J_1$ is small and the $\chi_{\bij}$ form a $[\pi,0]$ flux Ansatz, we can derive the effect of SSC interactions of the spectrum.
The mean-field parameters $\chi_{\bij}$ in this context are all real, shown in Fig.~\ref{fig:chiralansatz}(a).
The mean-field hoppings in Eq.~\eqref{eq:H0chiraldecomp} will become complex. We write the Hamiltonian as
\begin{equation}
    \mathcal{H}_0 = - \sum_{\bi\bj} (t_{\bij} f^\dagger_{\bj\alpha} f^\dagga_{\bi\alpha} + \hc)
\end{equation}
and now aim to derive the hoppings $t_{\bij}$.
Focusing on a single triangle $\bi\bj\bk$, we can evaluate the $\bk\bi$ hopping
\begin{equation}\label{eq:tki_chiralmft}
  t_{\bk\bi} = \frac{J_1\chi}{2} \left [ 1 - \iu \left( \frac{3\chi J_\chi}{2J_1} \right) \right] .
\end{equation}
The magnitude of the hopping increases with the added interaction, and there is an additional phase from the time-reversal symmetry breaking contribution $J_\chi$.
We find for the other two bonds that they are exactly the negative $t_{\bij} = t_{\bj\bk} = -t_{\bk\bi}$.
Therefore, a spinon hopping around this plaquette picks up a phase $\phi$, equal to three times the phase of Eq.~\eqref{eq:tki_chiralmft}. 
Likewise, on the upwards facing triangle, the result is $-\phi$.
The parameters are
\begin{equation}\label{eq:chiraltphi}
  t = \frac{J_1\chi}{2} \sqrt{1+\left( \frac{3\chi J_\chi}{2J_1} \right)^2}\, ,
  \quad
  \phi = -3\arctan(\frac{3\chi J_\chi}{2J_1}) \, .
\end{equation}
This result shifts the mean-field state from $[\pi,0]$ flux to $[\pi-\phi, \phi]$, with a modified normalization of the Hamiltonian.
Performing a gauge transformation, we show that the state can be parameterized as shown in Supplementary Fig.~\ref{fig:chiralansatz}. The Hamiltonian in the sublattice basis reads
\begin{equation}
    H_{\alpha\beta}^{XY}(\bm{k})= t\,\delta_{\alpha\beta}\,M_{\bm{k}}^{XY}
\end{equation}
where explicitly the phases are given by
\begin{align}
    M^{AA}_{\bm{k}}&= -M^{BB}_{\bm{k}} = 2 \cos(k_2) \nonumber\\
    M^{AB}_{\bm{k}}&= 1 + \eu^{-\iu k_1} + \eu^{\iu (\phi+k_2-k_1)} - \eu^{-\iu (\phi+k_2)}\, .
    \label{eq:subl_blochham_chiral}
\end{align}
The case $\phi=0$ recovers the U(1) Dirac spin liquid dispersion with gapless spinon bands. For finite $\phi$ the bands have a gap, shown in Supplementary Fig.~\ref{fig:chiralgap}.
Note that this continuous evolution of the angle $\phi$ is also seen in variational calculations \cite{hu2016variational}.

\begin{figure}
\centering
\includegraphics[width=0.6\textwidth]{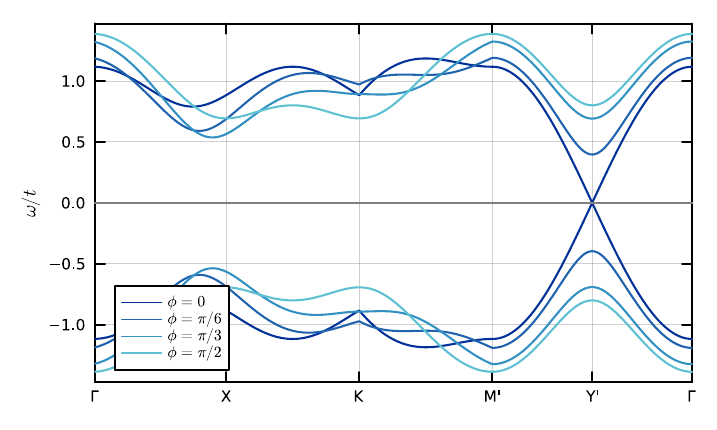}
\caption{{Spinon bands of the chiral spin liquid.} 
Chiral phases are as follows: gapless Dirac spin liquid Ansatz $\phi=0$; $\phi=\pi/6$; $\phi=\pi/3$; and translationally invariant time-reversal odd Ansatz $\phi=\pi/2$.
\label{fig:chiralgap}}
\end{figure}


\subsection*{Interacting susceptibility}
Now let us evaluate the dynamical response of the chiral QSL phase of this model.
We will evaluate the \emph{interacting} susceptibility of the fermionic spinons. 
The interactions will come from both $J_1$ and $J_2$ Heisenberg interactions, as well as the scalar spin-susceptibility.
The advantage of the self-consistent method defined above is that we may hope to produce quantitative predictions for the structure factor.

In this case, we are faced with treating a 6-fermion interaction.
In order to use the RPA, we have to make some further simplifications.
Look again at the form of the SSC interaction in the parton basis; the first term represents a down-spin hopping from $\bi$ to $\bj$, while an up-spin hops on the opposite path: $\bj$ to $\bk$ to $\bi$. The second term hops them in the opposite direction around the plaquette. These processes contribute with opposite phases.

The RPA sums over a series of bubble diagrams, where the interactions are four-fermion vertices. In the case of Heisenberg interactions, this was given by Eq.~\eqref{eqL:interaction_xybasis_sigma}.
In analogy, the SSC interaction takes the form
\begin{equation}
  J_{\chi} \sum_{\bi\bj\bj}  \vec{S}_{\bi}\cdot (\vec{S}_{\bj}\cross \vec{S}_{\bk})
= 
  J_\chi  \,  \frac{1}{8}\epsilon_{ijk} \sigma^i_{\alpha\beta}  \sigma^j_{\gamma\delta}\sigma^k_{\sigma\rho}  \sum_{\bi\bj\bk} 
f^\dagger_{\bi\alpha} f^\dagga_{\bi\beta} \, 
f^\dagger_{\bj\gamma} f^\dagga_{\bj\delta}\,
f^\dagger_{\bk\sigma} f^\dagga_{\bk\rho} \, .
    \label{eqL:interaction_xybasis_ssc}
\end{equation}
The RPA is based on a perturbative expansion of the interacting correlation function in terms of this interaction operator.
The contribution of this interaction at first order would give an antisymmetric, off-diagonal (in spin indices $ij$) contribution to $\chi(\bq,\omega)^{ij}_{XY}$. However, this contribution is proportional to various expectation values of spin operators $\langle f^\dagger_{\bi\sigma} \sigma_{\sigma\rho}^\dagga f^\dagga_{\bi\rho} \rangle$ and $\langle f^\dagger_{\bi\sigma} \sigma_{\sigma\rho}^\dagga f^\dagga_{\bj\rho} \rangle$.
We expect these contributions to all vanish in the mean-field state.
Therefore, the SSC interaction will only contribute at higher order.

The leading four-fermion interaction is therefore obtained from the second-order cumulant,
\begin{equation}
    H_{\mathrm{eff}}^{(2)}=-\frac{1}{2}\int d\tau\,\langle H_\chi(0)H_\chi(\tau)\rangle_{0,\mathrm{fast}}.
\end{equation}
Consider two counterclockwise-oriented triangles $(iab)$ and $(jba)$ sharing the bond $(ab)$. Contracting the spins on the common sites gives
\begin{equation}
    \int d\tau\,\langle S_a^b(0)S_a^{b'}(\tau)\rangle_0\langle S_b^c(0)S_b^{c'}(\tau)\rangle_0\simeq\frac{\delta^{bb'}\delta^{cc'}}{16J_1},
\end{equation}
where the correlation time of the virtual intermediate state has been approximated by $J_1^{-1}$. Since the shared bond appears with opposite orientation in the two triangles, $\epsilon^{abc}\epsilon^{a'cb}=-2\delta^{aa'}$ and hence
\begin{equation}
    \int d\tau\,\chi_{iab}(0)\chi_{jba}(\tau)\simeq-\frac{1}{8J_1}\,\mathbf S_i\cdot\mathbf S_j.
\end{equation}
The factor $1/2$ in the cumulant is canceled by the two time/orderings of each pair of adjacent triangles. Summing over all pairs of triangles sharing an edge therefore yields
\begin{equation}
    H_{\mathrm{eff}}\simeq\frac{J_\chi^2}{8J_1}\sum_{ij\in\mathcal B}\mathbf S_i\cdot\mathbf S_j,
\end{equation}
where $\mathcal B$ contains the nearest- and next-nearest-neighbor pairs generated by the corresponding triangle pairs. 
We confirm \textit{a posteriori} that the effect of this interaction is small, justifying the extension of the perturbative treatment up to large $J_\chi/J_1=0.5$. The dominant effect of increasing the SSC interaction is to increase the gap and stabilise the QSL.

\bibliographystyle{unsrt}
\bibliography{main}